\RequirePackage{ifpdf}
\documentclass[12pt,letterpaper]{JHEP3}         
\usepackage{amssymb,amsmath}
\usepackage{amsfonts,amsthm}
\usepackage{feynmp-auto}

\usepackage{wrapfig}
\usepackage{epsfig}

\def\be{\begin{eqnarray}}
\def\ee{\end{eqnarray}}
\newcommand{\nn}{\nonumber}
\newcommand\para{\paragraph{}}

\newcommand{\eqn}[1]{(\ref{#1})}

\def\Dslash{\,\,{\raise.15ex\hbox{/}\mkern-12mu D}}
\def\Dbarslash{\,\,{\raise.15ex\hbox{/}\mkern-12mu {\bar D}}}
\def\delslash{\,\,{\raise.15ex\hbox{/}\mkern-9mu \partial}}
\def\delbarslash{\,\,{\raise.15ex\hbox{/}\mkern-9mu {\bar\partial}}}
\def\pslash{\,\,{\raise.15ex\hbox{/}\mkern-9mu p}}
\def\calDslash{\,\,{\raise.15ex\hbox{/}\mkern-12mu {\cal D}}}

\def\implies{\Rightarrow}

\def\lae{\mathrel{\mathop{\smash{\lower .5 ex \hbox{$\stackrel<\sim$}}}}}
\def\lae{\mathrel{\mathop{\smash{\lower .5 ex \hbox{$\stackrel>\sim$}}}}}

\DeclareMathOperator{\Tr}{Tr}

\makeatletter
\newcommand*{\letterdef@}{}
\newcommand*{\letterdef}[3]{%
  \def\letterdef@##1{\expandafter\newcommand\csname #1\endcsname{#2{##1}}}%
  \@tfor\@tempa :=#3\do{\expandafter\letterdef@\expandafter{\@tempa}}}
\makeatother
\letterdef{bb#1}{\mathbb} {CHPRZ}    
\letterdef{c#1}{\mathcal}{ABCDEFGHIJKLMNOPQRSTUVWXYZ} 
\letterdef{rm#1}{\mathrm} {dDF} 
\letterdef{bf#1}{\mathbf}{pqkxyz}

\title{On Superconformal Anyons}

\author{Nima Doroud, David Tong and Carl Turner\\
Department of Applied Mathematics and Theoretical Physics, \\
University of Cambridge, \\ 
Cambridge, CB3 OWA, UK \\
{\tt  ndoroud, d.tong, c.p.turner@damtp.cam.ac.uk}\\}

\abstract{In $d=2+1$ dimensions, there exist field theories which are non-relativistic and superconformal. These theories describe two species of anyons, whose spins differ by $1/2$, interacting in a harmonic trap. We compute the dimensions of chiral primary operators. These operators receive large anomalous dimensions which are related to the unusual  angular momentum properties of anyons. Surprisingly, we find that the dimensions of some chiral primary operators violate the unitarity bound and we trace this to the fact that the associated wavefunctions become non-normalisable. We also study  BPS non-perturbative states in this theory: these are Jackiw-Pi vortices. We show that these emerge at exactly the point where perturbative operators hit the unitarity bound.  To describe the low-energy dynamics of these vortices, we construct a novel type of supersymmetric gauged linear sigma model.}

\begin{document}
\pagestyle{plain} \setcounter{page}{1}
\newcounter{bean}
\baselineskip16pt \setcounter{section}{0}
\unitlength = 1mm


\section{Introduction and Summary} 

It's hard to be free when you're an anyon. Statistical obligations mean that you constantly have to be aware of your standing relative to your neighbours. 
This has consequences. Even the simplest quantum mechanics problem involving multiple anyons is challenging. For example, the spectrum of anyons in a harmonic trap remains unsolved. The purpose of this paper is to study this simple system, and its supersymmetric extension, from the perspective of non-relativistic conformal field theory.

\para
It was realised long ago that the dynamics of anyons has a hidden $SO(2,1)$ conformal symmetry. This is true  whether the problem is expressed in the ``first-quantised" formalism of quantum mechanics \cite{roman}, or in the ``second-quantised" formalism of field theory \cite{jpi}. In the latter, field theoretic framework, anyons are described in terms of non-relativistic matter fields interacting with a Chern-Simons gauge field. 

\para
More recently, Nishida and Son revisited the quantum dynamics of non-relativistic field theories which exhibit  conformal invariance \cite{son}. They  showed, among other things, that there is a non-relativistic version of the state-operator map, with the spectrum of the dilatation operator mapped to the  spectrum of the Hamiltonian in the presence of a harmonic trap.
In this manner, solving the  quantum mechanics problem of anyons in a trap is equivalent to finding the scaling dimensions of operators in a non-relativistic conformal field theory.

\para
In this paper we will exploit the fact that one may introduce {\it superconformal} symmetry into the problem. Indeed, a non-relativistic superconformal field theory of anyons was constructed in \cite{llm}. It consists of two species of anyons whose spin differs by $1/2$. If we focus on the sector of the Hilbert space where only a single species of anyon is populated, this reduces to the problem above\footnote{This is a property of non-relativistic theories which is not shared by the more familiar relativistic supersymmetric theories. The bosonic sector of a relativistic theory knows about the presence of fermions through loop effects. But in a non-relativistic theory there are no anti-particles and, correspondingly, no virtual loops involving anti-particles.}.

\para
Embedding the problem in a larger theory with superconformal symmetry will turn out to be useful in exhibiting some of the properties of the anyonic spectrum. The representation theory of the non-relativistic superconformal group was discussed in \cite{nakayama,3lee}. There are both long multiplets and short  multiplets. The short multiplets are built on (anti)-chiral primary operators whose dimension $\Delta$ is dictated by the algebra
\be \Delta = \pm \left( J - \frac{3}{2} R\right)\label{done}\ee
where $J$ is the angular momentum of the operator, while $R$ is an R-symmetry, counting (roughly) the difference between the population of the two species of anyons. The spectrum of chiral primary operators has been discussed in both \cite{nakayama} and \cite{3lee}. However, there are a number of subtleties that arise in this spectrum that were previously overlooked. 

\para
The first subtlety is associated to the angular momentum of multiple anyons. Suppose that a single anyonic particle has spin $j$. If $n$ identical anyons are placed in a trap, their combined spin has the unusual property of scaling quadratically with the number of particles \cite{karabali,gold,djt}:   $J=n^2j$. This well known result follows directly from the spin-statistics theorem: exchanging two groups of $n$ anyons requires each anyon from the first group to be exchanged with each from the second, giving the $n^2$ behaviour to the statistical phase. The spin-statistics theorem, which is valid as our theory is the low-energy limit of a relativistic theory, then says that this scaling  should  also be manifest in the spin.

\para
Here, our interest lies in the implications of this result for \eqn{done} since it means that the dimension of $n$-particle operators should also scale as $n^2$. Indeed, we will show that the spectrum of $n$-particle chiral primary operators is given by
\be\Delta = n - n(n-1)j\label{result}\ee
The first term is the classical dimension of the operator. We will see, following \cite{son}, that the second term arises as an anomalous dimension for chiral primary operators. This anomalous dimension is one-loop exact. 

\para
The exact result \eqn{result} gives rise to a new puzzle because for $j>0$ the dimension of the operator would appear to become arbitrarily negative for a sufficiently large number of particles. This is in conflict with unitarity which requires $\Delta \geq 1$. We resolve this puzzle. We will see that the $j>0$ spectrum corresponds to anyons interacting through an attractive delta-function potential. This potential means that the quantum mechanical wavefunction diverges as two anyons approach. For a small number of anyons this is not an issue. However, if enough anyons are present this divergence means that the wavefunction ceases to be normalisable. The violation of the unitarity bound is due to this non-normalisability of the wavefunction. 

\para
The upshot of this analysis is that if too many anyons are placed in the same trap then the na\"ive ground state, in which each pair of anyons sits in the s-wave, is not normalisable. Good wavefunctions can only be constructed by giving pairs of anyons further relative orbital angular momentum.

\para
The story above is told in Sections  \ref{scftsec}  and \ref{spectrumsec}, and the appendix. Section \ref{scftsec} reviews the general structure of non-relativistic superconformal theories, including the state-operator map and the properties of chiral-primary operators. Section \ref{spectrumsec} derives the spectrum \eqn{result} of chiral anyons and discusses the unitarity bound and its relationship to  quantum-mechanical wavefunctions. The appendix contains a number of relevant one-loop computations.

\para
In Section \ref{jpvortexsec}, we change tack slightly. We describe non-perturbative configurations in the theory known as  Jackiw-Pi vortices \cite{jpvort}. We show that, rather remarkably, these solitons first appear at exactly the point in the spectrum where bound states of anyons first hit the unitarity bound. It is natural to  conjecture that these solitons can be viewed as a bound state of multiple anyons which now move as a single particle although, as we discuss, this proposal is not without its difficulties. 
Finally, we construct a novel supersymmetric quantum mechanics, in the form of a gauged linear sigma-model, to describe the low-energy dynamics of these vortices.

\section{Superconformal Field Theory}\label{scftsec}

Starting from Chern-Simons theories coupled to gapped, relativistic matter, one may take a non-relativistic limit in which anti-particles decouple but particles remain. Surprisingly, supersymmetry not only survives this limit but is enhanced to a non-relativistic superconformal (or super-Schr\"odinger) symmetry. The first construction of this type was presented in \cite{llm} for an ${\cal N}=2$ Abelian Chern-Simons theory coupled to a single chiral multiplet. 
Subsequent generalisations to other gauge groups, and different amounts of supersymmetry, were described in \cite{3lee,nak1,nak2,nak3,murugan}.

\para
In this paper we discuss the resulting non-relativistic physics. We will work with a $U(N_c)$ Chern-Simons theory, coupled to $N_f$ complex scalars $\phi_i$ and $N_f$ complex Grassmann-valued fields $\psi_i$, each of which transforms in the fundamental representation of the gauge group. The action is\footnote{Conventions: the subscripts $\mu,\nu,\rho=0,1,2$ run over both space and time indices, while $a=1,2$ runs over spatial indices only and $i=1,\ldots,N_f$ labels the flavour. The covariant derivatives read ${\cal D}_\mu \phi =\partial_\mu \phi - i A_\mu \phi$ and similarly for $\psi$. The magnetic field is $B= \partial_1 A_2-\partial_2A_1-i[A_1,A_2]$. We will later work with complex coordinates on the plane,  $z=x^1+ix^2$ and $\bar{z}= x^1-ix^2$ and the corresponding derivatives $\partial_z =\frac{1}{2}(\partial_1-i\partial_2)$ and $\partial_{\bar{z}}=\frac{1}{2}(\partial_1+i\partial_2)$.}
\be S = \int dt d^2x  && \left\{  i\phi_i^\dagger {\cal D}_0 \phi_i + i\psi_i^\dagger {\cal D}_0\psi_i -  \frac{k}{4\pi} {\rm Tr}\,\epsilon^{\mu\nu\rho}(A_\mu \partial_\nu A_\rho - \frac{2i}{3} A_\mu A_\nu A_\rho) \right.  \nn\\ &&\ \ \ \ \  \ \ \ 
 - \frac{1}{2m}\left( {\cal D}_a \phi_i^\dagger\, {\cal D}_a \phi_i + {\cal D}_a \psi_i^\dagger\,{\cal D}_a \psi_i -\psi_i^\dagger B\psi_i\right)  \label{lag} \\&&\ \ \ \ \  \ \ \  \left. - \frac{\pi}{mk}\left( (\phi_j^\dagger\phi_i)(\phi_i^\dagger\phi_j) - (\phi_j^\dagger\psi_i)(\psi_i^\dagger\phi_j) + 2 (\phi_i^\dagger\phi_j)(\psi_j^\dagger \psi_i) \right) \right\}\ \ \ \ \ \ \ \nn \ee
Both $\phi$ and $\psi$ fields give rise to excitations with mass $m$. The first order kinetic terms mean that there are particles, but no anti-particles. The $\phi$ excitations have spin $-1/2k$, while the $\psi$ excitations have spin $1/2 - 1/2k$. In what follows we will sometimes refer to the $\phi$ excitations as ``bosons" and the $\psi$ excitations as ``fermions", despite the fact that both are in general anyonic. The action \eqn{lag} also admits a relevant deformation consistent with supersymmetry which, as discussed in \cite{susyqhe}, drives the system to a quantum Hall state. We will not turn on this deformation here. 

\para
The quartic potential terms describe a delta-function contact interaction between anyons. The potential between bosons depends on the sign\footnote{There is a second class of theories, related to \eqn{lag} by a parity transformation. This flips the sign of the Chern-Simons term and the Pauli term, leaving the quartic potential invariant.} of $k$: it is repulsive for $k>0$ and attractive for $k<0$. This distinction will play an important role in this paper and the physics of anyons is rather different in the two cases. 

\para
The Chern-Simons coupling in \eqn{lag} ensures that any particle excitation is accompanied by magnetic flux. This is imposed through Gauss' law  which, classically,  reads
\be B = \frac{2\pi}{k}\left( \phi_i\phi_i^\dagger - \psi_i\psi_i^\dagger  \right)\label{gauss}\ee
At the quantum level, this means that operators $\phi_i$ and $\psi_i$ must be dressed with flux to make them gauge invariant. This is easily achieved in the $N_c=1$ Abelian theory as explained, for example, in \cite{3lee}. We introduce the  dual photon $\sigma$, normalised to have periodicity $2\pi$. The Chern-Simons term ensures that $e^{i\sigma}$ has charge $-k$ under the $U(1)$ gauge symmetry and composite, gauge invariant operators can be defined as
\be \Phi_i = e^{-i\sigma/k} \phi_i\ \ \ ,\ \ \ \Psi_i=e^{-i\sigma/k}\psi_i\label{dress}\ee
For the non-Abelian $U(N_c)$ theories, it is more difficult to write explicit expressions for the local gauge invariant operators. (One can easily construct non-local operators through the addition of a Wilson line.) Nonetheless, one expects that such objects exist, transforming in the ${\bf N}_c$  representation of the global part of the gauge group (and in the ${\bf N}_f$ representation of the flavour group).

\para
At this point, we pause to note that the operators \eqn{dress} involve fractional monopole operators. These are acceptable in our non-relativistic theory, and also in gapped relativistic theories.  They would not, however, be allowed in a relativistic Chern-Simons matter theory at the conformal point\footnote{We thank Kimyeong Lee for a discussion on this point.}. (See, for example,  \cite{seok,seok1,is,ofer}.) One way to see this is to note that, in relativistic theories, the state-operator map takes monopole operators inserted in the plane to magnetic flux over the two-sphere, where standard Dirac quantisation ensures that it is integer valued. In contrast, in non-relativistic theories the operators are transformed to flux in a harmonic trap, where there is no such quantisation requirement. The upshot is that in conformal relativistic theories, one must work with operators of the  schematic form $e^{-i\sigma} \phi^k$. For us, however, $\Phi$ and $\Psi$ defined in \eqn{dress} are  sensible operators.

\subsection{Symmetries}\label{symmetrysec}

The action \eqn{lag} is invariant under a superconformal symmetry, both at the classical level \cite{llm} and at the quantum level \cite{bergman,bbak}.  Here we review the generators of this symmetry. We work in the Schr\"odinger picture so that,  in contrast to \cite{jpi,llm}, the operators below do not have any explicit time dependence.

\subsubsection*{Bosonic Symmetries}

The number of bosons and fermions in the theory are individually conserved, with Noether charges
\be {\cal N}_B = \int d^2x\  \rho_B \ \ \ {\rm and}\ \ \  {\cal N}_F = \int d^2x\ \rho_F\label{nbnf}\ee
with $\rho_B=\phi_i^\dagger\phi_i$ and $\rho_F = \psi_i^\dagger\psi_i$. 
We denote the total particle number as  ${\cal N} = {\cal N}_B + {\cal N}_F$. 
The difference between ${\cal N}_B$ and ${\cal N}_F$, suitably weighted, will become the R-symmetry of the theory. We postpone a discussion of the appropriate weighting to later. 

\para
The spacetime symmetries of the theory give rise to a number of conserved charges. 
The Hamiltonian $H$ can be most simply written as 
\be H = \frac{2}{m} \int d^2x \ \, {\cal D}_{\bar{z}}\phi_i^\dagger {\cal D}_z\phi_i  + {\cal D}_z\psi_i^\dagger {\cal D}_{\bar{z}}\psi_i +\frac{\pi}{k} (\phi_i^\dagger\phi_j)(\psi_j^\dagger\psi_i) \label{h}\ee
which coincides with the Legendre transform of the Lagrangian \eqn{lag} only after imposing Gauss' law \eqn{gauss}. 

\para
Translational invariance gives rise to momentum conservation. The conserved charge is naturally complexified as $P = \frac{1}{2}(P_1-iP_2)$, defined by
\be 
P = \int d^2x\ {\cal P}\ \ \ {\rm with}\ \ \ {\cal P} = -\frac{i}{2}\left(\phi_i^\dagger {\cal D}_z\phi_i - ({\cal D}_z\phi_i^\dagger)\phi_i + \psi_i^\dagger{\cal D}_z\psi_i - ({\cal D}_z\psi_i^\dagger)\psi_i\right)\ \ \ \label{p}\ee
The Lagrangian enjoys an invariance under Galilean boosts. These too are  naturally complexified as $G=\frac{1}{2}(G_1- iG_2)$ defined by
\be G = \frac{m}{2}\int d^2x\ \bar{z}(\rho_B+\rho_F)\label{g}\ee
The theory is also invariant under conformal transformations, generated by the dilatation operator
\be D = \int d^2x\ (z{\cal P} + \bar{z}\bar{\cal P})\label{d}\ee
and special conformal transformations
\be C = \frac{m}{2}\int d^2x\ |z|^2 (\rho_B+\rho_F)\label{c}\ee
Note that the values of $G$, $D$ and $C$ are not conserved under time evolution. However, they evolve in a simple manner so that it is straightforward to construct conserved quantities $\tilde{G} = tP - G$, $\tilde{D} = Ht-D$ and $\tilde{K} = -t^2 H + 2tD +K$.

\para

We quantize the fields with canonical equal time (anti)-commutation relations.
%
%
A well-known subtlety arises because of Gauss' law, which means that the gauge field does not commute with the matter fields \cite{jpi}. In the appendix, we describe the Feynman rules for this theory and exhibit a number of one-loop computations, including a review of the result of \cite{bergman,bbak} showing that conformal invariance persists at the quantum level.

\para
It is straightforward to check that the bosonic symmetry generators above obey the algebra \cite{jpi,llm}
\be i[D,P] = -P\ \ \ ,\ \ \ i[D,G] = +G\ \ \ ,\ \ \ i[D,H] = -2H\ \ \ ,\ \ \ i[D,C] = +2C\ \ \ \ \nn\\  i[C,P]=-G \ \ \ , \ \  \ 
 [H,G] = - i P\ \ \ ,\ \ \ [H,C]=-iD \ \ \ ,\ \ \ [P,G^\dagger]=-\frac{im}{2}{\cal N}\ \ \ \ 
\label{alg1}\ee
 with all remaining commutators vanishing. 
This is the Schr\"odinger algebra. The triplet of operators $H$, $D$ and $C$ form an $SO(2,1)$ subgroup.

\subsubsection*{Angular Momentum} 

We didn't include the angular momentum in the generators above because it deserves special attention. As usual for particles with internal spin, there are two contributions. The first comes from the orbital angular momentum and is given by
\be  J_0 = \int d^2x\ \left( i z{\cal P} - i\bar{z}\bar{\cal P} \right)
 \label{j0}\ee
This should be supplemented with the angular momentum due to the spin of the fields. Here there is some arbitrariness in the choice. We pick
\be J =  J_0 - \frac{1}{2k}{\cal N}_B + \left(\frac{1}{2}-\frac{1}{2k}\right){\cal N}_F \label{angmom}\ee
Note that this differs from the choice of \cite{llm,nakayama,3lee} by the presence of the terms proportional to $1/2k$, but as these are proportional to the total particle number ${\cal N}$ they do not change the structure of the algebra. Nor do they change the dimension of chiral primary operators defined through \eqn{done} because, as we shall see shortly, the R-symmetry picks up a compensating shift. However, the choice above will prove more convenient as it treats the spin of the boson and fermion on equal footing, so that
\be [J,\phi_i^\dagger] = -\frac{1}{2k}\phi_i^\dagger\ \ \ {\rm and}\ \ \ [J,\psi_i^\dagger] = \frac{k-1}{2k}\psi_i^\dagger\label{1pang}\ee
The same commutation relations also hold for the gauge invariant excitations $\Phi_i$ and $\Psi_i$ defined in \eqn{dress}.
The angular momentum operator has only two non-vanishing commutators with other bosonic symmetry generators:  $ [J,P]=-P$ and $[J,G]=-G$. 

\para
As we mentioned in the introduction, the  angular momentum of $n$ anyons has the unusual property, first discovered in \cite{karabali, gold}, that it is proportional to $n^2$ rather than $n$. This fact will play an important role in our analysis, so we pause here to review the underlying physics. More details can be found, for example, in the book \cite{lerda}.

\para
To demonstrate the addition of angular momentum, it will suffice to work classically. (For a derivation in the quantum theory see, for example, \cite{djt}.) We can illustrate this in the Abelian theory with $N_c=N_f=1$, focussing on the derivation for the bosonic field. The important term is the gauge field buried in the covariant derivatives in the expression for the orbital angular momentum \eqn{j0}. Picking a configuration with no traditional orbital angular momentum, we're still left with
\be J_0 =  -\int d^2x\ \epsilon^{ab}x_a A_b\,\rho_B\nn\ee
The gauge field is determined by Gauss' law \eqn{gauss}. Choosing the gauge $\partial_a A^a=0$, we can solve \eqn{gauss} for the vector field, giving 
\be A^a({\bf x})  = -\frac{2\pi}{k}\epsilon^{ab}\partial_b\int d^2x'\,G({\bf x}-{\bf x}')\rho_B({\bf x}')\label{thisisa}\ee
with $G({\bf x}-{\bf x}') = \frac{1}{2\pi}\log|{\bf x}-{\bf x}'|$ being the usual Green's function for the Laplacian in the plane. This gives the following contribution to the angular momentum:
\be J_0 = -\frac{2\pi}{k} \int d^2x\,d^2x'\,\rho({\bf x})\rho({\bf x}')x^a \partial_a G({\bf x}-{\bf x}')\nn\ee
We take the charge distribution to be a sum of delta-functions at $n$ distinct points ${\bf r}_i$,
\be \rho_B = \sum_{i=1}^n\delta^2({\bf x}_i - {\bf r}_i(t))\nn\ee
and the orbital angular momentum becomes
\be J_0 = -\frac{2\pi}{k}\sum_{i,j} {\bf r}_i\cdot\frac{\partial}{\partial {\bf r}_i}G({\bf r}_i-{\bf r}_j)\nn\ee
At this point we need a procedure to deal with the fact that this expression is ill-defined when ${\bf r}_i={\bf r}_j$. Any regularisation which preserves anti-symmetry under reflection gives $\lim_{x\rightarrow x'} \partial_a G({\bf x}-{\bf x}') = 0$. With this choice, the sum is over pairs of particles only and we have
\be J_0 = -\frac{2\pi}{k} \sum_{i=1}^n\sum_{j\neq i} {\bf r}_i\cdot\frac{\partial}{\partial {\bf r}_i} G({\bf r}_i-{\bf r}_j) = -\frac{n(n-1)}{2k}\nn\ee
This is the orbital angular momentum of $n$ anyons. To this, we must add the total spin which is $-n/2k$. The final result is 
\be J = -\frac{n^2}{2k}\label{addspin}\ee
as promised. Note that, in general, this is not the lowest angular momentum of $n$ anyons: in certain cases, one can decrease the spin by giving the individual particles additional  relative orbital angular momentum. The result \eqn{addspin} is, however, the  angular momentum  that one gets when adiabatically increasing the statistical parameter of $n$ bosons. 

\subsubsection*{Fermionic Symmetries}

The action \eqn{lag} enjoys a  number of fermionic symmetries \cite{llm}. There are two complex supersymmetries, generated by 
\be q = i\sqrt{\frac{m}{2}}\int d^2x \ \phi_i^\dagger \psi_i\ \ \ {\rm and}\ \ \ Q =  \sqrt{\frac{2}{m}}\int d^2x\ \phi_i^\dagger {\cal D}_{\bar{z}}\psi_i\label{q}\ee
 together with a superconformal symmetry, generated by
 \be S = i\sqrt{\frac{m}{2}}\int d^2x\ z\phi_i^\dagger\psi_i\label{s}\ee
These operators all have spin $\pm 1/2$, as seen from the commutation relations
\be [J,q]= \frac{1}{2}q\ \  , \ \ [J,Q] = -\frac{1}{2}Q\ \ ,\ \ [J,S] = -\frac{1}{2}S\label{jcom}\ee 
The anti-commutators in the fermionic symmetry generators give the algebra 
\be &\{q,q^\dagger\}& \!= \frac{m}{2}{\cal N}\ \ \ ,\ \ \ \{Q,Q^\dagger\} = H\ \ \ , \ \ \ \{q,Q^\dagger\} =  P \label{alg2}\\ &\{S,S^\dagger\}&=C\ \ \ ,\ \ \ \{q,S^\dagger\}= -G\ \ \ ,\ \ \ \{Q,S^\dagger\} = \frac{i}{2}\left(iD-J_0+{\cal N}_B-{\cal N}_F\right)\nn\ee
The last of these is important. We rewrite it in terms of the angular momentum operator \eqn{angmom}, including the spin contribution, as 
\be \{Q,S^\dagger\} = \frac{i}{2}\left(iD - J + \frac{3}{2}R\right)\label{qs}\ee
This defines the R-symmetry, which is given by
\be
 R = \frac{2k-1}{3k}{\cal N}_B - \frac{k+1}{3k}{\cal N}_F \label{r}\ee
 The R-symmetry of each of the fermionic symmetry generators is
 \be [R,q]=-q\ \ \ ,\ \ \ [R,Q]=-Q\ \ \ ,\ \ \ [R,S]=-S\label{rcom}\ee 
 The remaining non-vanishing commutators between bosonic and fermionic generators are
 \be& i[D,Q]=-Q\ \ \ ,\ \ \ i[D,S] = S&\nn\\
& i[C,Q]=S\ \ \ ,\ \ \ i[H,S] =-Q\ \ \,\ \ i[P,S]=i[G,Q]=-q&\label{alg3}\ee
 The various (anti-)commutators collected in \eqn{alg1}, \eqn{jcom}, \eqn{alg2}, \eqn{rcom} and \eqn{alg3} make up the non-relativistic ${\cal N}=2$ superconformal algebra, also known as the super-Schr\"odinger algebra. These algebras have a long history; some early papers include \cite{ggt,llm,duval}.
 
 \subsection{Operators, States and Chiral Primaries}\label{nicesec}
 
 In the remainder of this section we describe the structure of theories that are invariant under the super-Schr\"odinger algebra, following \cite{son,nakayama,3lee}. We will mostly discuss generalities, postponing a more detailed discussion of the theory \eqn{lag} to Section \ref{spectrumsec}.

 \para
 Our interest lies in the spectrum of the dilatation operator $D$. Consider the local operators of the theory ${\cal O}({\bf x}=0)$, evaluated at the origin. An operator is said to have {\it scaling dimension} $\Delta_{\cal O}$ if it obeys
\be i[D,{\cal O}] = -\Delta_{\cal O} {\cal O}\nn\ee
Since $D$ commutes with $J$, ${\cal N}$ and $R$ we can simultaneously assign angular momentum $j_{\cal O}$, particle number $N_{\cal O}$, and R-charge $r_{\cal O}$ to operators ${\cal O}$ with fixed dimension, 
\be [J,{\cal O}] = j_{\cal O}{\cal O}\ \ \ ,\ \ \ [{\cal N},{\cal O}] = N_{\cal O}{\cal O}\ \  \ ,\ \ \ [R,{\cal O}] = r_{\cal O} {\cal O}\nn\ee
Given an operator ${\cal O}$ of dimension $\Delta_{\cal O}$, one can construct further operators of fixed dimension using the algebra \eqn{alg1}. 
\be i[D,[H,{\cal O}]] &=& -(\Delta_{\cal O}+2)[H,{\cal O}]\nn\\  
i[D,[P_a,{\cal O}]] &=& -(\Delta_{\cal O}+1)[P_{a},{\cal O}]\nn\\  
i[D,[G_a,{\cal O}]] &=& -(\Delta_{\cal O}-1)[G_{a},{\cal O}]\nn\\
i[D,[C,{\cal O}]] &=& -(\Delta_{\cal O}-2)[C,{\cal O}]\nn\ee
 We see that both the boosts $G_a$ and the special conformal transformation $C$ lower the dimension of the operator. Assuming that the spectrum of $D$ is bounded below, this must end somewhere. The places where it ends are called {\it primary operators} \cite{son}. (See also \cite{henkel}.) These are operators of fixed scaling dimension that also obey
 \be [G_a,{\cal O}] = [C,{\cal O}] = 0\nn\ee
 We can then construct an infinite tower of operators, starting from the primary and acting with $P_a$ and $H$. These are {\it descendants}. The tower forms an irreducible representation of the Schr\"odinger algebra.

\para
From the commutators of $D$ with the fermionic generators \eqn{alg3}, we see that $S$ lowers the dimension of an operator, and $Q$ raises the dimension, while $q$ leaves the dimension unchanged. We define a {\it superconformal primary} by the further requirement that 
\be [S,{\cal O}] = [S^\dagger,{\cal O}]=0\label{primark}\ee
where the commutators are replaced with anti-commutators if ${\cal O}$ itself is fermionic. Finally, there are two further special classes of operators that will be of particular importance \cite{nakayama,3lee}. These are {\it chiral primary} operators  which, in addition to \eqn{primark}, obey
\be [Q, {\cal O}]=0\nn\ee
and {\it anti-chiral primary} operators which, in addition to \eqn{primark}, obey
\be [Q^\dagger,{\cal O}]=0\nn\ee
We'll see the importance of these below.

 \subsubsection*{Introducing a Harmonic Trap: the State-Operator Map}

In relativistic theories, the state-operator map equates the spectrum of the dilatation operator on the plane to the spectrum of the Hamiltonian of the theory defined on a sphere. In the non-relativistic context, there is also a state-operator map. Its interpretation is arguably even more physical: the spectrum of the dilatation operator is equal to the spectrum of the Hamiltonian when the theory is placed in a harmonic trap \cite{son}. 

\para
The map can be thought of as a field-theoretic generalisation of the usual approach to conformal quantum mechanics \cite{aff}. We introduce the new Hamiltonian
\be L_0 = H + C\label{l0}\ee
From the definition of the special conformal generator \eqn{c}, we see that this is indeed equivalent to placing the theory in a harmonic trap.

\para
For each  primary operator ${\cal O}$, we define the state
\be |\Psi_{\cal O}\rangle = e^{-H} {\cal O}(0) |0\rangle\label{map}\ee
where $|0\rangle$ is the vacuum of the original Hamiltonian $H$, the state with no particles present.  It is simple to check, using the algebra \eqn{alg1}, that 
\be L_0 |\Psi_{\cal O}\rangle = \Delta_{\cal O}|\Psi_{\cal O}\rangle\label{state}\ee
This is the promised result: the spectrum of $L_0$ acting on states coincides with the spectrum of $D$ acting on operators. 
The quantum numbers $j$, $N$ and $r$ which label the operator also carry over to label the state, so we have $J |\Psi_{\cal O}\rangle = j_{\cal O} |\Psi_{\cal O}\rangle$, ${\cal N} |\Psi_{\cal O}\rangle=N_{\cal O} |\Psi_{\cal O}\rangle$ and $R |\Psi_{\cal O}\rangle=r_{\cal O} |\Psi_{\cal O}\rangle$.

\para
In the presence of the trap, it is useful to redefine the remaining operators. We introduce 
\be L_\pm = H - C \pm iD\ \ \ ,\ \ \ {\cal P}_a = P_a + iG_a\ \ \ ,\ \ \ {\cal G}_a = P_a - iG_a\nn\ee
(We previously used complex notation to denote momentum \eqn{p} and boosts \eqn{g}; this complex structure is different from the one introduced above so we revert to spatial coordinates here.) These new operators obey
\be [L_0,L_\pm] = \pm 2 L_\pm\ \ ,\ \ [L_+,L_-] = -4 L_0\ \ ,\ \ [L_0,{\cal P}_a] = {\cal P}_a\ \ ,\ \ [L_0,{\cal G}_a] = -{\cal G}_a 
 \nn\ee
together with a number of other non-trivial commutation relations which follow from \eqn{alg1}. We see that $L_+$ and ${\cal P}_a$ increase the energy of the state by $2$ and $1$ respectively. Meanwhile, $L_-$ and ${\cal G}_a$ decrease the energy, shifting it by $2$ and $1$ respectively. It is simple to show that primary states of the form \eqn{state}, built from primary operators, obey 
$L_- |\Psi_{\cal O}\rangle = {\cal G}_a |\Psi_{\cal O}\rangle =0$. These lie at the bottom of the tower. 
One can then construct higher energy states by acting with either $L_+$ or  with ${\cal P}_a$. These are descendant states. Some of these descendant states are null \cite{nakayama}.

\subsubsection*{Chiral Primary States}

The super-Schr\"odinger algebra places further structure on the theory. Both operators and states sit in supersymmetric multiplets. Here we retain the harmonic trap and  describe the states of the Hamiltonian $L_0$. It is simple to adapt the supercharges to this set-up \cite{3lee}. We define
\begin{figure}[!h]
\begin{center}
\includegraphics[height=1.0in]{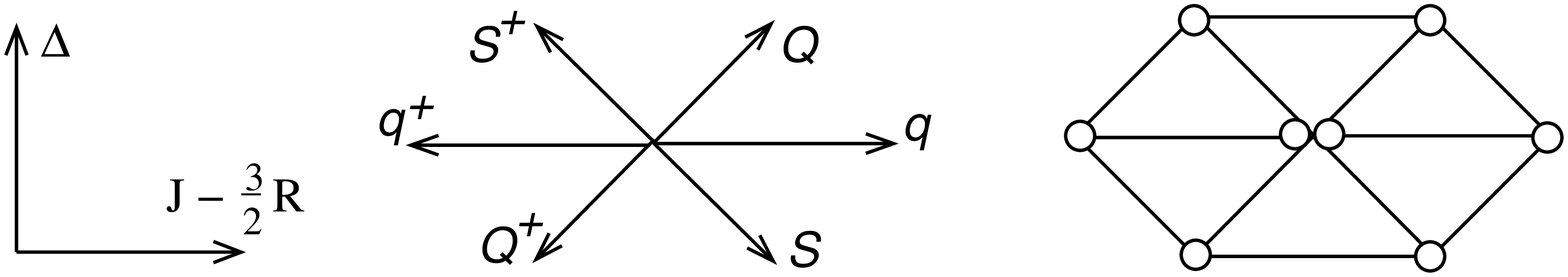}
\end{center}
\caption{A generic supersymmetric multiplet, following \cite{3lee}.}\label{multifig}
\end{figure}
\noindent
\be {\cal Q} = Q - iS\ \ \ {\rm and}\ \ \ {\cal S} = Q +  iS\label{qands}\ee
These obey the algebra
\be \{{\cal Q},{\cal Q}^\dagger\} = L_0 + \left( J-\frac{3}{2}R\right)\ \ &,&\ \  \{ {\cal Q},{\cal S}^\dagger\} =  L_+ 
\label{qsnew}\\
\{{\cal S},{\cal S}^\dagger\} = L_0 - \left( J - \frac{3}{2}R\right)\ \ \ &,& \ \ \{{\cal Q}^\dagger,{\cal S}\} = L_-\nn\ee
Their commutators with the generators $L_0$ and $L_\pm$ are given by
\be [L_0,{\cal Q}] = {\cal Q}\ \ \ ,\ \ \ [L_0,{\cal Q}^\dagger] = -{\cal Q}^\dagger\ \ \ &,&\ \ \ [L_0,{\cal S}] = -{\cal S}\ \ \ ,\ \ \ [L_0,{\cal S}^\dagger] = {\cal S}^\dagger
\nn\\{}\  [L_+,{\cal Q}] = [L_{-},{\cal S}]=0\ \  \ ,\ \ \ [L_-\!&,&\!{\cal Q}]= 2{\cal S}\ \ \ ,\ \ \ [L_{+},{\cal S}] = -2{\cal Q}
\label{lqs}\ee
We see that, acting on an eigenstate of $L_0$, the operators ${\cal Q}$ and ${\cal S}^\dagger$ raise the energy, while ${\cal Q}^\dagger$ and ${\cal S}$ lower the energy. The upshot is that a superconformal primary operator gives rise to a superconformal primary state,  sitting at the bottom of a tower and obeying
\be L_-|\Psi_{\cal O}\rangle = {\cal Q}^\dagger|\Psi_{\cal O}\rangle = {\cal S}|\Psi_{\cal O}\rangle=0\label{scprimary}\ee
Representations of the super-Schr\"odinger algebra sit in supersymmetric multiplets, built on these superconformal primary states \cite{nakayama,3lee}. There is a unique trivial multiplet: the 
vacuum state, which is annihilated by all supercharges and, in our theory, has quantum numbers $\Delta = N = j=r=0$.

\para 
A generic excited state sits in a long multiplet. This contains 8 primary states. The action of the supercharges $q$, ${\cal Q}$ and ${\cal S}$ on these states is, following \cite{3lee}, shown in the figure.

\para
There are also short multiplets in which the dimension of the superconformal primary is fixed by the superconformal algebra. These are the states that interest us here. A chiral primary  operator gives rise to a chiral primary state obeying, in addition to \eqn{scprimary}, 
\be [Q,{\cal O}]  = 0 \ \ \ \Leftrightarrow\ \ \  {\cal Q}|\Psi_{\cal O}\rangle  =0\nn\ee
\EPSFIGURE{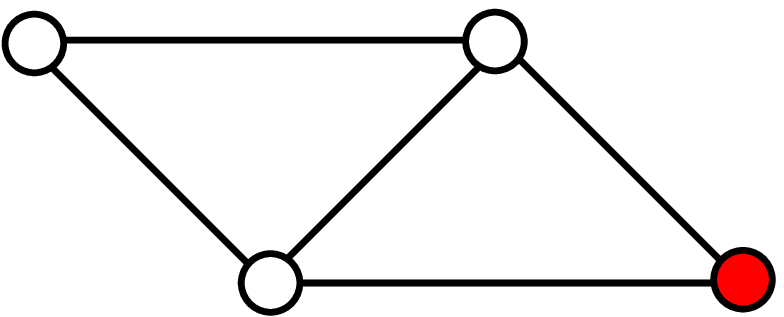,height=39pt}{A chiral multiplet}
\noindent
 The associated multiplet contains four primary states,  as shown in the figure. Of these, one is special, denoted by the red dot; its quantum numbers are dictated by the algebra \eqn{qsnew} and satisfy
\be \Delta_{\cal O} = -\left( j_{\cal O} -\frac{3}{2}r_{\cal O}\right)\label{cp}\ee
In our theory, the operators $\phi^\dagger_i$ and $\psi_i$ are chiral primary. Or, if we insist on gauge invariance, the dressed operators $\Phi^\dagger_i$ and $\Psi_i$ defined in \eqn{dress} are chiral primary. However, only creation operators give rise to physical states, which means that the only chiral primary states are those involving excitations of $\Phi^\dagger$. We will describe these states in more detail in the next section. 

\para
An anti-chiral primary operator gives rise to an anti-chiral primary state which obeys, in addition to \eqn{scprimary}, 
\be [Q^\dagger,{\cal O}] =0\ \ \ \Leftrightarrow\ \ \ {\cal S}^\dagger|\Psi_{\cal O}\rangle  =0\nn\ee
\EPSFIGURE{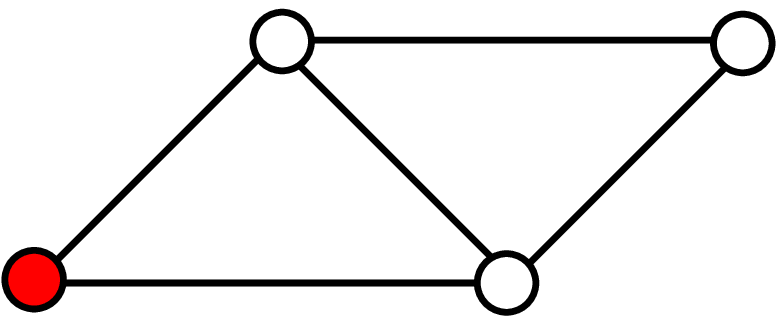,height=40pt}{An anti-chiral multiplet}
\noindent
(Note that the shift from $Q^\dagger$ on operators to ${\cal S}^\dagger$ acting on states arises due to the factor of $e^{-H}$ in the state-operator map \eqn{map} and the commutation relations \eqn{alg3}.)
There are again four primary states in the multiplet, as shown in the figure. One of these, denoted by the red dot, obeys
\be \Delta_{\cal O} = +\left( j_{\cal O} -\frac{3}{2}r_{\cal O}\right)\label{acp}\ee
The operators $\Phi_i$ and $\Psi_i^\dagger$ are anti-chiral primary. The anti-chiral primary states contain  $\Psi_i^\dagger$ excitations only. 


\subsubsection*{Unitarity Bounds}

There are a number of constraints on the quantum numbers that follow from the algebra alone. The first of these holds for any non-relativistic conformal theory and is particularly simple
\be \left|\left| \, {\cal P}_a|\Psi_{\cal O}\rangle\,\right|\right|^2 \geq 0 \ \ \ \Rightarrow\ \ \ N_{\cal O} \geq 0\nn\ee
This is the statement that there are no anti-particles in the theory. The second  constraint also holds for any non-relativistic conformal theory. Assuming that we are not in the vacuum state, so $N_{\cal O}\neq 0$, we have \cite{nagain}
\be \left|\left| \, 2m{\cal N}L_+ - {\cal P}_a{\cal P}_a|\Psi_{\cal O}\rangle\,\right|\right|^2  \geq 0\ \ \ \Rightarrow\ \ \ \Delta_{\cal O}\geq 1\label{bound1}\ee
Further, states that saturate this bound obey the equation
\be L_+\,|\Psi_{\cal O}\rangle = \frac{1}{2mN_{\cal O}}{\cal P}_a{\cal P}_a\,|\Psi_{\cal O}\rangle\label{free}\ee
This looks, formally, like the  Schr\"odinger equation for a free particle. (Recall that in relativistic theories, saturation of a unitarity bound indicates that the operator is free.)

\para
The final bound follows from the superconformal invariance of the theory. We look at the combination
\be \left.\begin{array}{c} \langle\Psi_{\cal O}|\{{\cal Q},{\cal Q}^\dagger\}|\Psi_{\cal O}\rangle \geq 0 \nn\\ \langle\Psi_{\cal O}|\{{\cal S},{\cal S}^\dagger\}|\Psi_{\cal O}\rangle \geq 0 \end{array}\ \ \ \right\}  \ \ \ \Rightarrow \ \ \ \Delta_{\cal O} \geq \left| j_{\cal O} - \frac{3}{2}r_{\cal O}\right|\nn\ee
Clearly this bound is saturated by the (anti)-chiral primary states.

\section{The Spectrum}\label{spectrumsec}

In this section we describe the spectrum of (anti)-chiral primary states in the field theory defined by \eqn{lag}. The physics is rather different depending on whether the interactions are repulsive ($k>0$) or attractive ($k<0$) and we treat the two cases separately. Moreover, the key ideas already arise for the simplest theory with $N_c=N_f=1$ and we concentrate on this case, mentioning the generalisation to non-Abelian theories with more flavours only in passing.

\subsection{Repulsive Interactions}

We start by describing the situation with $k>0$. In this case, the quartic terms in \eqn{lag} describe repulsive delta-function contact interactions between the anyons. This ensures that the wavefunction vanishes as anyons coincide \cite{bergman,son}, a point of view that we will explain in some detail in Section \ref{qmsec}.

\para
The chiral primary states in our theory arise from excitations of the gauge-invariant composite bosonic operator $\Phi^\dagger$ and the momentum $P$.  The $n$-particle chiral primary state of lowest dimension is 
\be {\cal O} = (\Phi^\dagger)^n\nn\ee
The charge of this operator under the R-symmetry \eqn{r} scales linearly with $n$. However, 
as explained in the previous section, the angular momentum of this operator does not: it is instead given by \eqn{addspin},
\be j_{\cal O} = -\frac{n^2}{2k}\ \ \ ,\ \ \ r_{\cal O} = \left(\frac{2k-1}{3k}\right)n\label{jnr}\ee
The chiral primary constraint \eqn{cp} then fixes the dimension of this operator to be
\be \Delta_{\cal O} = n  + \frac{n(n-1)}{2k}\label{cpdim}\ee
It is unusual for the dimension of an operator to scale as $n^2$. The classical dimension of $(\Phi^\dagger)^n$ is simply $n$. This means that the second term above, proportional to $1/2k$, must arise as an anomalous dimension. This is indeed the case. This was shown already for the Abelian theory in \cite{son}. In Appendix \ref{appa}, we extend this result to the non-Abelian theory.  The chiral primary bound ensures that the dimension is one-loop exact.

\para
The states with energy \eqn{cpdim} and their descendants correspond to a class of known, exact solutions to the quantum mechanics of $n$ anyons \cite{chou,murthy}. (Reviews of the anyon spectrum can be found,  for example,  in \cite{khare,date}.)
These are sometimes called ``linear" solutions in the literature because their energy scales linearly with the statistical parameter $1/2k$. The structure of the superconformal theory could, somewhat generously, be said to explain the existence of this class of solutions. Unfortunately, the superconformal formulation does not seem to   help with the so-called ``non-linear solutions"; these correspond to long multiplets which are unprotected by the chiral primary bound and receive corrections at two loops and, presumably, higher. 

\def\chirald{\partial_{z}}
\def\achirald{\partial_{\bar z}}

\para
Primary operators with higher dimension can be constructed via the insertion of the momentum operators $P$ and $P^\dagger$ or, equivalently, the derivatives $\partial_z$ and $\partial_{\bar z}$. Those that consist of $\chirald$ only are chiral primary since each insertion of $\chirald$ decreases the angular momentum by 1, while increasing the dimension by 1. 
 As an  example, consider the $n=2$ particle sector. The operators with one derivative are all descendants. There are six  operators with two derivatives: $\Phi^\dagger \chirald^2\Phi^\dagger$, $\Phi^\dagger\achirald^2\Phi^\dagger$, $\Phi^\dagger\chirald\achirald \Phi^\dagger $, $\achirald\Phi^\dagger \achirald\Phi^\dagger $, $\chirald\Phi^\dagger\chirald \Phi^\dagger $ and $\achirald\Phi^\dagger \chirald\Phi^\dagger$. Of these, three linear combinations are total derivatives and hence descendants; the remaining three are primary.  One of these primary states is chiral primary.

\para
We note that, for $k\gg 1$,  the chiral primary $(\Phi^\dagger)^n$ is the $n$ particle ground state. However, it is known from the study of anyons that, when $n\geq 3$, it is not the true ground state for all $k$. Instead, there is a level crossing (as $k$ is varied) and, for small $k$,  a long multiplet becomes the ground state. For the Abelian theory, it is found numerically that this cross-over happens around  $k\approx 1.4$ for $n=3$  and $k\approx 1.8$ for $n=4$ \cite{verb0,verb}. 

\para
The anti-chiral primary states arise from the composite fermionic operator $\Psi^\dagger$ and the momentum $P^\dagger$. However, when $N_{f}=N_{c}=1$ the Grassmann nature of $\Psi$ forbids us from constructing operators of the form $(\Psi^\dagger)^n$.
 In the $n$-particle sector, the anti-chiral primary operator of the lowest dimension is instead
\be \tilde{\cal O} = \Psi^\dagger \achirald\Psi^\dagger \ldots \achirald^{n-1} \Psi^\dagger \label{otilde}\ee
A single fermionic excitation $\Psi^\dagger$ has angular momentum $1/2-1/2k$. The operator $\tilde{\cal O}$ has angular momentum
\be j_{\tilde{\cal O}} = \left(\frac{1}{2}-\frac{1}{2k}\right)n^2 = \frac{n}{2} + \frac{n(n-1)}{2} - \frac{n^2}{2k} \nn\ee
The first way of writing $j_{\tilde{\cal O}}$ mimics the expression for the bosonic angular momentum \eqn{jnr}; the terms in the second expression can be thought of respectively as the angular momentum of $n$ fermions, the angular momentum induced by the derivatives and the angular momentum due to the flux attachment. The R-charge of $\tilde{\cal O}$ is $ r_{\tilde{\cal O}} = -(k+1)n/3k$.  The superconformal algebra \eqn{acp} then ensures that the dimension of these anti-chiral primary operators is given by 
\be \Delta_{\tilde{\cal O}} = \frac{n(n+1)}{2} - \frac{n(n-1)}{2k}\label{acpdim}\ee
Again, the first term is the classical dimension of the operator $\tilde{\cal O}$; the second term arises as a one-loop anomalous dimension as we show in Appendix \ref{appa}.

\para
One can easily lower the classical dimension of the operator $\tilde{\cal O}$ by including both $\partial_z$ and $\partial_{\bar z}$ in the string, resulting in an operator with classical dimension $\sim n^{3/2}$ rather than $n^2$. This operator is primary but not chiral: it gives rise to a long multiplet of states that, for $k\gg 1$, is the ground state in this sector. However, as $k$ decreases there is a level crossing and the anti-chiral primary operator $\tilde{\cal O}$ becomes the ground state.

\para
The bosonic chiral primary operators \eqn{cpdim} and the fermionic  anti-chiral primary operators \eqn{acpdim} actually trace out the same spectrum as $k$ is varied continuously: they are related by $1/k\rightarrow 1 - 1/k$. Indeed, at $k=1$, the $\Phi$ excitations are fermions and the $\Psi$ excitations are bosons and we have
\be {\cal O}_{k=1} = \tilde{\cal O}_{k=\infty}\ \ \ {\rm and}\ \ \ \tilde{\cal O}_{k=1} = {\cal O}_{k=\infty}\nn\ee
However, in the non-Abelian theory $k$ is necessarily quantised and the spectra of ${\cal O}$ and $\tilde{\cal O}$ do not coincide as $k$ is varied. Moreover, when $N_c>1$ or $N_f>1$, the representations of the bosonic and fermionic operators under the global symmetries differ as described in Appendix \ref{appa}. 

\para
It may be interesting to explore this spectrum further to look for remnants of the dualities that occur in relativistic Chern-Simons theories, such as \cite{gs,min1,ofer1,min2}. 

\subsection{Attractive Interactions: the View from Quantum Mechanics}\label{qmsec}

We now turn to the case of $k<0$. This describes an attractive delta-function contact interaction between anyons and, as we will see, brings  a  surprise. This is because the expressions for the angular momentum, R-charge and, ultimately the dimensions of (anti)-chiral primary operators are the same as before.

\para
For the fermionic anti-chiral primary states, there is nothing to worry about. In contrast, for $k<0$, the bosonic chiral primary states have energy
\be \Delta_{\cal O} = n - \frac{n(n-1)}{2|k|}\label{strange}\ee
For a sufficiently large number of particles, this contradicts the unitarity bound $\Delta \geq 1$! We hit the bound when $n=2|k|$ and violate the bound when $n>2|k|$. This requires an explanation. What's going on?

\subsubsection*{Quantum Mechanics of Anyons}

To understand why these states violate the unitarity bound, we turn to the quantum mechanical description of the problem. Such a formulation exists because there are no anti-particles in the Lagrangian \eqn{lag} and, moreover, the dynamics of the gauge field is tied to that of the particles. This means that there can be no fluctuation of particle number so, if  you fix the number of  bosons and fermions in the problem, then the field theory \eqn{lag} reduces to the quantum mechanics of a finite number of degrees of freedom. A derivation of how to move from the field theory language to the quantum mechanics can be found, for example, in the book \cite{lerda}.

\para
Here we consider the sector with $n$ ``bosons" and no ``fermions". Each particle has position $x^a_i$, with $a=1,2$ the spatial index and  $i=1,\ldots,n$ labelling the particle.  The quantum mechanics Hamiltonian is 
\be
H = - \frac 1{2m}\sum_{i=1}^{n} \left(\partial_a^{\,i} + \frac{i}{k} \epsilon_{ab} \,\partial_b^{\,j}\, \sum_{j\neq i} \log|{\bf x}_i - {\bf x}_j| \right)^2  +\frac{2\pi}{mk} \,\sum_{i<j} \delta^2({\bf x}_i - {\bf x}_j) \ \ \ \ \ \label{qmham}
\ee
Here the log term arises from the gauge field which is given by \eqn{thisisa}; this is the term which imposes anyonic statistics on the particles which now pick up a phase $\pi/k$ when exchanged. The delta-functions arise from the $|\phi|^4$ interactions in the Lagrangian. These contact interactions are repulsive for $k>0$ and attractive for $k<0$. We should ultimately add to the Hamiltonian \eqn{qmham} the harmonic potential. We'll do this below, but it won't be important for our immediate discussion.

\para
For us, the role played by the delta-function contact interactions is key. These arise naturally from the field theory and endow the quantum mechanics with a number of nice features. Indeed, as we review below, they are necessary for the quantum mechanics to exhibit scale invariance. For now, their main purpose is to impose boundary conditions on the  wavefunction\footnote{In this section, we will use $\Psi$ to denote the wavefunction of bosonic anyons. It is not to be confused with the composite operator introduced in \eqn{dress} to denote fermionic anyons.}
 $\Psi({\bf x}_i)$ as anyons get close to each other. For two particles, their s-wave state has boundary condition  
\be \Psi({\bf x}_1,{\bf x}_2) \sim |{\bf x}_1-{\bf x}_2|^{1/k}\ \ \ \ {\rm as}\ \ {\bf x}_1\rightarrow {\bf x}_2\label{wfbc}\ee
with a pairwise generalisation to multiple particles\footnote{There is an alternative way to view these boundary conditions. One could  exclude from the configuration space the points where particles meet and propose a self-adjoint extension of the quantum mechanics \cite{manuel}. There are precisely two such extensions which are compatible with scale invariance, corresponding to \eqn{wfbc} with $k>0$ and $k<0$.}.

\para
For repulsive contact interactions \emph{i.e.} $k>0$, the wavefunction \eqn{wfbc}  vanishes as the particles approach; it is equivalent to imposing a hard-core boundary condition.

\para
In contrast, with an attractive contact interaction, corresponding to $k<0$, the wavefunction diverges as the two particles approach. For two particles, this is not problematic because the wavefunction \eqn{wfbc} is normalisable as long as $|k|>1$. But this divergence becomes more serious when we add too many particles, as we now describe.

\para
The wavefunction for $n$ particles in which each pair sits in the s-wave is
\be \Psi_0 = \prod_{i<j} | {\bf x}_i - {\bf x}_j |^{-1/|k|} \label{psi0}\ee
This corresponds to the operator $(\Phi^\dagger)^n$. (We will make the connection between operators and wavefunctions more precise below.) One can check that $\Psi_0$ is a zero-energy eigenstate of the Hamiltonian \eqn{qmham}. As discussed in the next section, there is actually a large degeneracy at zero-energy, so one might reasonably ask what is special about this eigenstate. The answer is that it is the one adiabatically connected to the ground state \eqn{psi0tilde} in the presence of a trap breaking this degeneracy. (The connection to $(\Phi^\dagger)^n$ makes this no surprise.)

\para
When all $n$ particles coincide, one finds a divergence from each of the $n(n-1)/2$ pairs of particles. This means that the wavefunction takes the schematic form $\Psi_0\sim {r^{-n(n-1)/2|k|}}$
 where $r$ measures the ``radial" relative distance from the coincident point. The normalisation is 
 \be \int \prod_{i=1}^nd^2x_i \ |\Psi_0|^2 \sim \int d^2X\int dr\,r^{2n-3} |\Psi_0|^2 \sim \int d^2X \int dr\, \frac{r^{2n-3}}{r^{n(n-1)/|k|}} \label{norm}\ee
where $X$ is the centre of mass. We see that the norm is UV finite if and only if
\be 2n- 3 - \frac{n(n-1)}{|k|} > -1 \qquad \iff \qquad n < 2|k| \nn\ee
The wavefunction is normalisable only when $ n < 2|k|$. This, of course, coincides with the threshold that we found from the unitarity bound \eqn{strange}. 

\para
This, then, is the answer to our puzzle: the operators $(\Phi^\dagger)^n$ which violate the unitarity bound correspond to wavefunctions in the quantum mechanics which are non-normalisable. Note, in particular, that the wavefunction with $n=2|k|$ particles is also (logarithmically) non-normalisable, despite the fact that the operator saturates the unitarity bound. The relationship between violations of the unitarity bound and the non-normalisability of the wavefunction was previously noted in a different context in \cite{nishy}.

\subsubsection*{Mapping Between Operators and Wavefunctions}

We have learnt that the operator $(\Phi^\dagger)^n$ corresponds to a non-normalisable state for $n\geq 2|k|$. This leaves open the simple question: what is the ground state of $n\geq 2|k|$ anyons in a trap? In general, there is no reason to believe that the ground state lies in a chiral multiplet. This makes the question difficult. We can, however, answer the simpler question: what is the lowest energy chiral state for $n\geq 2|k|$?

\para
To answer this question, we will extend the correspondence
\be {\cal O} = (\Phi^\dagger)^n \ \ \ \longleftrightarrow \ \ \ \Psi_0=\prod_{i<j} |{\bf x}_i-{\bf x}_j|^{1/k}\nn\ee
to other chiral operators. This will allow us to determine which operators ${\cal O}$ correspond to normalisable wavefunctions.

\para
To proceed, we rewrite our Hamiltonian \eqn{qmham} in a way which eliminates the delta-functions. Indeed, there is independently good reason to do this, related to the scale invariance of the theory. In two dimensions, the delta-function has the same scaling as the Laplacian $\nabla^2$, which means that  a Hamiltonian of the form \eqn{qmham} would appear to be scale invariant for any coefficient of the delta-function interaction. This is misleading. Delta-functions in quantum mechanics require a regularisation and this typically breaks scale invariance, resulting in a simple example of an anomaly in a quantum mechanical setting \cite{beg,jbeg,holstein}. A similar effect also arises from the log term in \eqn{qmham}. For the choice of coefficient in front of the delta-function in \eqn{qmham}, these two effects cancel.  This, of course, mimics the field theory analysis of \cite{bergman,bbak}.

\para
To see the cancellation explicitly, we can rewrite the Hamiltonian by introducing complex coordinates $z_i=x_i^1+ix_i^2$ for each particle. For the specific coefficient of the delta-function interaction given in \eqn{qmham}, we have
\be
H = - \frac{2}{m} \sum_{i=1}^{n} \Psi_0 \left( \, \partial_{\bar{z}_i} \partial_{z_i} + \frac 1 k  \sum_{j\neq i} \frac {\partial_{z_i}}{\bar{z}_i - \bar{z}_j} \right) \Psi_0^{-1}  \label{complexham}
\ee
where $\Psi_0$ is given in \eqn{psi0}.  The $\partial$ operators in \eqn{complexham} are understood to act on everything to the right including, ultimately, the wavefunction.

\para
This form of the Hamiltonian \eqn{complexham} has no delta-functions and, correspondingly, no need for regularisation: it provides a manifestly scale invariant description of the dynamics. Further, it is immediately clear that the wavefunction $\Psi_0$ obeys $H\Psi_0=0$ as previously claimed.
It is also  easy to write down a large class  of  eigenfunctions, given by
\be \Psi = \bar{f}(\bar{z}_1, \ldots, \bar{z}_n) \prod_{i<j} | z_i - z_j |^{1/k} \label{chiralwf}\ee
where $\bar{f}(\bar{z})$ is an antiholomorphic function, symmetric in its arguments $\bar{z}_i$.
We propose that this class of wavefunctions is equivalent to the set of chiral operators of the conformal field theory, with the mapping given up to normalisation by
\be  {\cal O} = \chirald^{p_1} \Phi^\dagger \cdots \chirald^{p_n} \Phi^\dagger \qquad \longleftrightarrow \qquad f = z_1^{p_1} \cdots z_n^{p_n} + \mbox{permutations}\nn \ee
In particular we see that descendants in the CFT, which are obtained by total derivatives, correspond to choices of $f$ with factors of $\sum z_i$:
\be  \tilde{{\cal O}} = \chirald {\cal O} \qquad \longleftrightarrow \qquad \tilde{f} = \left(\sum_{i=1}^n z_i \right) f \nn \ee
Hence, if we exclude the descendants we are left to form $f$ by symmetrising products of terms like $(z_i-z_j)^2$ which respect the bosonic properties of the particles. This provides a useful way to describe and enumerate all chiral primaries. For example, there are no chiral primaries with just a single derivative while, at the  two derivative level, we find
\be  {\cal O} = \left(\chirald^2 \Phi^\dagger \Phi^\dagger - \chirald \Phi^\dagger \chirald \Phi^\dagger \right) \Phi^\dagger \cdots \Phi^\dagger \qquad \longleftrightarrow \qquad f = (z_1 - z_2)^2 + \mbox{other pairs} \nn \ee
In this manner, we see that chiral primary operators arise by giving pairs of particles extra relative angular momentum. Correspondingly, the angular momentum of chiral primaries is spaced in even-integer steps, since we always need polynomials of even order in $z$. In contrast,  the angular momentum of descendants is spaced in integer steps.

\para
Now we can ask  which of these wavefunctions  lie in the Hilbert space. The divergences of the chiral wavefunctions \eqn{chiralwf} are softer than those of $\Psi_0$. Heuristically, this is because the addition of $(z_i-z_j)^2$ factors increases the relative angular momentum of a pair of particles, so that their wavefunction is damped where they meet. However, we need to determine  what form of $f$ is sufficient to render the wavefunctions normalisable as the particles converge.

\para
Let us suppose that $f$ is a polynomial of order $2m$. The simplest criterion on the wavefunction arises from the situation where all $n$ particles converge to a point. In this case, repeating the calculation \eqn{norm},  the requirement for normalisability is
\be 2n - 3- \frac{n(n-1)}{|k|} + 4m >  -1 \qquad \iff \qquad 2m > (n-1)\left(\frac n {2|k|} - 1 \right)  \nn \ee
This coincides with the requirement that the dimension of the corresponding operator, which is schematically of the form $\mathcal{O} \sim \chirald^{2m} (\Phi^\dagger)^n$, sits strictly above the unitarity bound  $\Delta >  1$. This agreement is reassuring but it is not the end of the story.

\para
Suppose that we instead bring some subset of $q<n$ particles together. Without loss of generality, we can pick particles $i=1,\ldots,q$. The wavefunction \eqn{chiralwf} diverges as $r^{2m_q -  q(q-1)/2|k|}$ where $m_q$ is the {\it smallest} number of relative angular momentum terms $(z_i - z_j)^2$ with $i,j=1,\ldots,q$ that appears in the expansion of $f$.  Clearly when we include all particles we include all winding terms, so $m_n = m$.

\para
This is perhaps best illustrated with an example. Consider  $n=4$, with $f \sim (z_1 - z_2)^2(z_1 - z_3)^2 + \cdots$. We see that $m_4 = 2$ is the total number of angular momentum terms. However,  $m_3 = 0$ because $f$ remains of  order 1 if particle 1 is  separated while particles $2$, $3$ and $4$ are brought together.  Thus the additional angular momentum in $f$ has not helped convergence at $q=3$.

\para
The significance of this is that there are additional constraints at each order $q$ on the form of $f$ and, correspondingly, on the possible chiral operators ${\cal O}$ in the theory. These constraints are equivalent to imposing that $\Delta > 1$ not only for the operator $\mathcal O$ itself, but for every `factor' of $\mathcal O$: that is to say, if we can express ${\cal O} = {\cal O}_1 {\cal O}_2$ then we need $\Delta_{{\cal O}_i} > 1$ as well.

\para
Nonetheless, by including enough angular momentum, one may see that it is in fact always possible to find UV-normalisable chiral states in the theory.  The relative angular momentum forms a barrier, supporting the wavefunction away from the origin so that the wavefunction survives in the Hilbert space.

\subsubsection*{Solutions in the Trap}

Finally, for completeness we observe that we can also find explicit chiral wavefunctions in a trap. The Hamiltonian is now
\be L_0 = H + \frac{m}{2}  \sum_i |{\bf x}_i|^2\nn\ee
It is again more convenient to express this in complex coordinates in a manner analogous to \eqn{complexham}. It is
\be L_0 = \tilde{\Psi}_0\left[ \sum_{i=1}^{n} \Bigg( - \frac{2}{m} \partial_{\bar{z}_i} - \frac 2 {mk}  \sum_{j\neq i} \frac {1}{\bar{z}_i - \bar{z}_j} + z_i   \Bigg) \partial_{z_i}  + n + \frac{n(n-1)}{2k} + \sum_{i=1}^n \bar{z}_i \partial_{\bar{z}_i} \right]\tilde{\Psi}_0^{-1} \nn \ee
where $\tilde{\Psi}_0$ is the ground state wavefunction in the trap,
\be \tilde{\Psi}_0 =   \prod_{i<j} |z_i - z_j|^{1/k}  \exp \left(-\frac{m}2\sum_{i=1}^n |z_i|^2 \right)\label{psi0tilde} \ee
Hence writing $\Psi = \bar{f}(\bar{z}) \tilde{\Psi}_0$ with $f$ any symmetric degree $d$ polynomial, we analytically find a class of wavefunctions with energies
\be
\Delta = n + \frac{n(n-1)}{2k} + d
\nn\ee
This coincides with \eqn{cpdim} when $d=0$ and with the general chiral bound \eqn{cp} for $d\neq 0$.

\section{Jackiw-Pi Vortices}\label{jpvortexsec}

This section has a somewhat different focus. The Lagrangian \eqn{lag} admits soliton solutions. These are non-topological vortices, first discovered by Jackiw and Pi \cite{jpvort,jpi}. 
Despite a vast literature on Jackiw-Pi vortices, their role in the quantum dynamics of anyons has not, to our knowledge, been explained. We do not present a full picture here. Instead we describe various aspects of their dynamics and point out a few surprising connections to aspects of anyons that we described above.

\subsection{Vortices in the Plane}

We start with a review of Jackiw-Pi vortices. (More detailed expositions can be found, for example, in \cite{dunne,horvathy}.) They are classical solitons which, in the $U(1)$ theory with $N_f=1$, obey the equations
\be B = \frac{2\pi}{k} |\phi|^2  \ \ \ , \ \ \ {\cal D}_z\phi=0\label{jpi}\ee
together with $\psi=0$. Solutions exist only in theories with $k<0$.  They have vanishing Hamiltonian $H=0$,  as defined in \eqn{h}.

\para
It is simple to solve \eqn{jpi}. We decompose the scalar field as
\be \phi = \sqrt{\rho}e^{i\chi}\label{phiisthis}\ee
where $\rho  = \phi^\dagger\phi$ is the matter density. The gauge field is determined by the second equation in \eqn{jpi}  to be
\be A_z = \partial_z\chi -\frac{i}{2}\partial_z\log \rho \label{thisisaz}\ee
Substituting this into the Gauss law constraint reveals that $\rho$ satisfies the Liouville equation, 
%
%
\be \nabla^2 \log\rho = \frac{4\pi}{k}\rho\nn\ee
The general solution for $k<0$ can be written in terms of a holomorphic function $f(z)$,
\be \rho = \frac{k}{2\pi} \nabla^2\log\left(1+|f(z)|^2\right)\label{jpsol}\ee
It is illuminating to look at the axially symmetric solutions, with $f(z) = (z_0/z)^p$. These take the form
\be \rho = \frac{2|k|p^2 r_0^{2p}}{\pi} \frac{ r^{2(p-1)}}{(r_0^{2p} + r^{2p})^2}\label{vortexrho}\ee
Asymptotically, the matter density scales as $\rho \sim r^{-2(p+1)}$ and normalisability requires that $p>0$. Meanwhile, at the origin, the matter density scales as $\rho \sim r^{2(p-1)}$. To ensure that the gauge field \eqn{thisisaz} is non-singular, the phase of $\chi$ must wind accordingly. This requires  $p$ to be integer, with the scalar field profile given by
\be \phi =   \sqrt{\frac{2|k|p^2r_0^{2p} }{\pi}}\, \frac{  r^{(p-1)}}{r_0^{2p} + r^{2p}} \,e^{-(p-1)\theta}\nn\ee
This means that, although there is no topology in the vacuum manifold supporting these solitions, their charge is nonetheless quantised. The integral of the matter density is
\be n = \int d^2x\ \rho = 2|k| p\label{jpiflux}\ee
and the corresponding flux is $\int B  = -4\pi p$. Note that the minimal flux carried by the vortices is  twice that required by flux quantisation.  This is a well known, if rather peculiar, feature of classical Jackiw-Pi vortices.

\para
Although we derived \eqn{jpiflux} for axially symmetric solutions, it continues to hold for the most general solution. For separated vortices, we may take
\be f(z) = \sum_{a=1}^p \frac{c_a}{z-z_a}\label{separate}\ee
This describes $p$ vortices at positions $z_a$, with $c_a$ providing a scale size and phase for each vortex. (This solution needs amending as the vortices coincide.) The collective coordinates $z_a$ and $c_a$ parameterise the moduli space ${\cal M}_p$ of Jackiw-Pi vortices which has dimension ${\rm dim}\,{\cal M}_p=4p$.

\para
There is something striking about the result \eqn{jpiflux}: setting $p=1$, we see the single vortex has $n=2|k|$, which is exactly the same point where the operator $(\Phi^\dagger)^n$ hits the unitarity bound $\Delta=1$! An operator at the unitarity bound should describe a single, free excitation. It is therefore natural to conjecture that semi-classically, a suitably regularised  $(\Phi^\dagger)^n$ operator creates a Jackiw-Pi vortex.

\para
Although this conjecture is natural, we have not been able to find corroborating evidence. For example, 
the orbital angular momentum, defined in \eqn{j0}, becomes $J_0=n = 2|k|p$ when evaluated on vortices. Note that this is linear in $p$, rather than quadratic. The full angular momentum \eqn{angmom}, which includes a contribution from the particle number, is then 
\be J= 2|k|p+p\nn\ee
This angular momentum is greater than that of any chiral primary operator. (Recall that $(\Phi^\dagger)^{2|k|}$ has $J = 2|k|$, while including $P$ within an operator decreases the angular momentum.)

\para
Moreover, it is not clear that we are comparing like for like. As we have discussed, when the theory is defined on the plane one should talk about the spectrum of $D$ acting on local operators. But it is difficult to interpret soliton solutions in terms of local operators. Instead, it is  conceptually simpler to think about the solitons as states in the theory, in which case it is appropriate to consider solitons in the  presence of a harmonic trap.

\subsection{Vortices in a Trap}

The Hamiltonian with a harmonic potential is given by
\be L_0 = H + \omega^2 C = \frac{2}{m} \int \rmd^2x\left( |{\cal D}_z\phi|^2 + \frac{m^2 \omega^2}{4} |z|^2 |\phi|^2 \right) \label{l0init}\ee
where we have introduced $\omega^2$, the strength of the trap. (In our previous discussion, we implicitly set $\omega = 1$.) First order equations for vortices in the trap can be derived by completing the square, with\footnote{There is a second way to complete the square in the Hamiltonian, resulting in the Bogomolnyi equation ${\cal D}_{\bar{z}} \phi = \pm i \frac{\omega m}{2}\bar{z}\phi$. Static solutions to this equation are directly related to the Jackiw-Pi solutions \eqn{jpsol} by the a spatially varying phase. This time the Hamiltonian is given by $L_0= \pm D$, the dilatation operator. This is not a conserved charge and the resulting time dependent configuration does not remain a solution to the  Bogomolnyi equation. These configurations are related, but not identical, to those discussed in  \cite{jpagain}.}
\be L_0 = \frac{2}{m} \int \rmd^2x\left( \left|{\cal D}_z\phi + \frac{\omega m}{2} \bar{z}\phi \right|^2 - \frac{\omega m}{2}\left( z\phi^\dagger{\cal D}_z\phi + \bar{z}{\cal D}_{\bar{z}}\phi^\dagger\phi\right)\right) \label{ursquare}\ee
The cross-terms are conserved charges, related to the orbital angular momentum given in \eqn{j0}. We can then write
\be L_0 \geq -\omega(J_0-{\cal N}_B)\nn\ee
Looking back at the definitions of the angular momentum and the R-symmetry, we see that this coincides with the  bound from the supersymmetry algebra \eqn{cp}. States which saturate the bound, with $L_0= -\omega(J_0-{\cal N}_B)$, are chiral primary states. They obey the Bogomolnyi equations
\be {\cal D}_z\phi = \frac{\omega m}{2}\bar{z}\phi\label{bog}\ee
Solving this equation gives a static configuration, independent of time. However, since the time evolution is governed by $L_0=-\omega(J_0-{\cal N}_B)$, both of which are conserved charges, the subsequent dynamics is straightforward: the soliton configuration rotates around the trap, together with an overall temporal phase. The energy of this solution is given by
\be L_0 = m\omega^2\int d^2x\ |z|^2 |\phi|^2\label{lisc}\ee
Solutions to \eqn{bog} describe semi-classical chiral states in the theory. It should be possible, in the appropriate regime, to match these onto the exact quantum states that we have constructed in Section \ref{spectrumsec}. Unfortunately, we do not currently understand enough about the solutions to \eqn{bog} to make any precise statements. Instead, we restrict ourselves to a few simple comments and leave a more detailed study of these solutions for future work.

\para
To solve the static equation \eqn{bog}, we again  decompose $\phi$ as \eqn{phiisthis} and   solve for the gauge field which now reads
\be A_z = -\frac{i\omega m}{2} \bar{z} + \partial_z\chi -\frac{i}{2}\partial_z\log \rho \nn\ee
Substituting this into Gauss' law, we get
\be \nabla^2\rho = \frac{4\pi}{k}\rho - 2\omega m\label{taubes}\ee
This equation no longer has analytically known solutions; indeed, the constant term means that it is similar to the Taubes equation \cite{taubes} which arises for relativistic BPS vortices, but with different minus signs. This constant term ensures that the matter density now falls off exponentially quickly due to the presence of the trap.

\para
An aside: usually when completing the square to derive Bogomolnyi equations, one has the option to pick $\pm$ signs corresponding to BPS or anti-BPS solitons. In contrast, in \eqn{ursquare} only one sign is allowed since the opposite sign gives rise to exponentially growing solutions rather than exponentially decaying. The Jackiw-Pi vortices therefore correspond to chiral states rather than anti-chiral states.

\para
The Taubes equation \eqn{taubes} is not analytically solvable. It is, however, well studied \cite{taubes}. It is known that a general solution with $\int B = -2\pi p $ has $2p$ collective coordinates. When $p$ is an even integer, this agrees with the number of moduli for the Jackiw-Pi vortices in the plane.  It would be interesting to understand the properties of this moduli space in more detail and explore to what extent these solutions can be viewed as the semi-classical version of the quantum chiral states.

\para
Here we make a few simple comments about the simplest class of axially symmetric solutions, with 
\be \phi = \sqrt{\rho(r)} e^{-iq\theta}\ \ \ \ \ q\in {\bf Z}\nn\ee
These axially symmetric solutions evolve in time only by an overall phase. We assume that the asymptotic form of the matter density is
\be \rho \sim \left\{\begin{array}{cc} Ce^{-\omega m r^2}\left(r^{2\lambda} + {\cal O}(r^{2(\lambda-1)})\right)\ \ \ & \ \ \ r\rightarrow \infty\\ r^{2q}\left(1+{\cal O}(r^2)\right)\ \ \  &\ \ \  r\rightarrow 0\end{array}\right.\nn\ee
We do not know which values of $q$ and $\lambda$ arise in solutions. There appears to be no integer restriction  on $\lambda$ through regularity of solutions alone. We can   manipulate the expressions for both particle number $n$ and energy $L_0$ into total derivatives to find that any solution of this form must have
\be n = \int d^2x\ |\phi|^2 = |k|(q-\lambda)\ \  \ {\rm and}\ \ \ L_0 = n(q+1) - \frac{n^2}{2|k|}\nn\ee
This provides certain restrictions on the allowed values of $q$ and $\lambda$ since we must have both $n>0$ and $L_0 >0$. 

\para
Numerically we can find solutions with $q=0$ for a range of $\lambda<0$. In particular, the solutions with $q=0$ and $\lambda=-1$ have $n=|k|$; this is half the particle number of the minimal Jackiw-Pi vortex on the plane. It seems that the striking coincidence that Jackiw-Pi vortices first appear where operators hit the unitarity bound holds only on the plane and does not extend to the harmonic trap. We do not currently understand the physics behind this.

\subsection{Dynamics of Vortices}\label{susyqmsec}

The low-energy dynamics of solitons can usually be described in terms of dynamics on the moduli space ${\cal M}_p$. Typically, such a description involves natural geometric quantities defined over ${\cal M}_p$ including metrics, various forms, vector bundles and potentials. However, the dynamics of Jackiw-Pi vortices differs from the more familiar relativistic solitons since the dynamics is first order and, in the absence of some external potential, the vortices do not go anywhere. On top of this, the action of supersymmetry on these vortices is also rather novel. 
For all these reasons, a full supersymmetric low-energy action for Jackiw-Pi vortices has not, to our knowledge, been previously constructed. 
(See, for example, \cite{baklee} for an early attempt to address some of these issues.) 

\para
We end this paper by constructing a supersymmetric quantum mechanics which, we propose, describes the low-energy dynamics of Jackiw-Pi vortices. The quantum mechanics is of the form of a gauged linear sigma model and is motivated, in part, by a similar construction for relativistic vortices \cite{htong}. We do not claim that our model captures all details of the dynamics.  However, it does capture all the symmetries of the problem which, as we shall see, have a non-trivial structure. We hope that it will prove useful in describing certain BPS properties of these vortices. At the very least, it offers a new class of supersymmetric gauged linear models.


\para
Let us first explain why the supersymmetry preserved by vortices has a slightly different form from usual. 
Vortices on the plane are annihilated by the supersymmetry generator $Q$ \eqn{q}, while vortices in the trap are annihilated by ${\cal Q}$ \eqn{qands}.  In the more familiar context of relativistic theories, supercharges which annihilate solitons descend to supersymmetries on their worldvolume. However, as discussed in \cite{susyqhe}, this is not what happens in our non-relativistic theory. The difference comes from the fact that the moduli space ${\cal M}_p$ plays the role of the {\it phase space} rather than the configuration space. Any supercharge, like $Q$ or ${\cal Q}$, which acts trivially at all points  in the phase space simply plays no role in the low-energy dynamics.
 
 \para
In contrast, both $q$ and $S$ (or ${\cal S}$) act non-trivially on the phase space and these two supercharges will be realised in the low-energy dynamics. Our goal is to construct a theory invariant under these two supersymmetries. Such a supersymmetric quantum mechanics cannot be of the usual type. This can be seen by looking at the form of the superconformal generator \eqn{s} which depends on the position in space and so, in the quantum mechanics, must depend on the position of the soliton. This means that the action of $S$ descends to a non-linearly realised supersymmetry on the soliton worldline. 

\para
Nonetheless, there is a very straightforward way to construct a class of supersymmetric quantum mechanics with the appropriate supersymmetries: one simply dimensionally reduces the class of theories \eqn{lag} that we started with. There are two supersymmetric multiplets. 
\begin{itemize}
\item Vector Multiplet: This contains a worldline gauge field $\alpha$, associated to a gauge group $G$, and a single complex scalar $Z$, transforming in the adjoint representation of $G$.
\item Matter Multiplet: This contains a complex scalar $\varphi$ and a complex Grassmann object $\chi$. Both transform in the same representation $R$ of the gauge group.
\end{itemize}
 %
%
%

\para
In what follows, we discuss the generalisation of the Jackiw-Pi vortices to the  non-Abelian  $U(N_c)$ gauge group with fundamental matter. The equations for vortices in a trap of strength $\omega$ now read
\be B = \frac{2\pi}{k}\sum_{i=1}^{N_f}\phi_i\phi_i^\dagger\ \ \ ,\ \ \ {\cal D}_z\phi_i=  \frac{\omega m}{2}\bar{z}\phi_i \label{jpeqn}\ee
To our knowledge, these non-Abelian solitons have not been previously studied. We consider non-Abelian Jackiw-Pi vortices in the theory with $N_f=N_c=N$. Obviously the case $N=1$ reduces to the usual Abelian Jackiw-Pi vortices. We propose that the dynamics of $p$ vortices, with $n=2|k|p$,  is described by a $U(p)$ gauged quantum mechanics, coupled to  $N$ matter  matter multiplets $(\varphi_i,\chi_i)$ in the fundamental representation and a further $N$ matter multiplets $(\tilde{\varphi}_i,\tilde{\chi}_i)$ in the anti-fundamental representation.
The action is
\be S = k\int dt\ &&\!\! \left\{i{\rm Tr}\,Z^\dagger{\cal D}_t Z  + i \sum_{i=1}^N \left[\varphi_i^\dagger {\cal D}_t\varphi_i + \chi^\dagger_i{\cal D}_t\chi_i + \tilde{\varphi}_i^\dagger {\cal D}_t\tilde{\varphi}_i
+  {\rm Tr}\,\tilde{\chi}_i^\dagger {\cal D}_t\tilde{\chi}_i\right]\right.\nn\\ &&\ \ \ \ - \sum_i\left[\varphi_i^\dagger Z^\dagger Z \varphi_i + \chi_i^\dagger Z Z^\dagger \chi_i  + \tilde{\varphi}_i Z Z^\dagger \tilde{\varphi}_i^\dagger - \tilde{\chi}_iZ^\dagger Z \tilde{\chi}_i^\dagger\right] \nn\\ &&\ \ \ \ \left. - \sum_{i,j} {\rm Tr}\,(\phi_i\chi_i^\dagger - \tilde{\chi}_i\tilde{\varphi}_i)(\chi_j\varphi_j^\dagger - \tilde{\varphi}_j^\dagger\tilde{\chi}_j) - {\rm Tr}\,Z^\dagger Z
 \right\} \label{matrixlag}\ee
Here ${\cal D}_tZ = \partial_t Z - i[\alpha,Z]$ and ${\cal D}_t\varphi = \partial\varphi - i\alpha\varphi$ and ${\cal D}_t\tilde{\varphi} = \partial_t\tilde{\varphi} + i \tilde{\varphi}\alpha$.
The kinetic term for the anti-fundamental fermion can also be written as $i{\rm Tr}\,\tilde{\chi}^\dagger {\cal D}_t\tilde{\chi} = -i ({\cal D}_t\tilde{\chi})\tilde{\chi}^\dagger$; the unusual minus sign ensures that we get a positive definite number operator for these excitations. 
The equation for the $U(p)$ gauge field $\alpha$ is Gauss' law. It tells us that
\be [Z,Z^\dagger] + \sum_{i=1}^N \Big[\varphi_i\varphi_i^\dagger  - \chi_i\chi_i^\dagger - \tilde{\varphi}_i^\dagger\tilde{\varphi}_i -\tilde{\chi}_i^\dagger\tilde{\chi}_i\Big]=0\label{another}\ee
The gauged quantum mechanics \eqn{matrixlag} describes dynamics on the space ${\cal M}$ defined by \eqn{another}, quotiented by the $U(p)$ gauge action. Our conjecture is that this space should be thought of as the moduli space of Jackiw-Pi vortices in the plane. 
The positions of the $p$ vortices are encoded in the eigenvalues of $Z$, at least when the solitons are well-separated. The scale size, phase and $U(N)$ orientation are contained in the gauge invariant bilinears $\tilde{\varphi}_i\varphi_j$.

\para
The moduli space ${\cal M}$ of the gauge theory admits a dilatation symmetry under which all fields, including $Z$, are scaled. This property is shared by the moduli space of Jackiw-Pi vortices in the plane. However, the dilatation symmetry is broken by the potential,  written  on the second and third lines of \eqn{matrixlag}. This potential plays the role of the Hamiltonian of the theory; we call it ${\mathbb L}_0$ 
\be {\mathbb L}_0 &=&    \varphi_i^\dagger Z^\dagger Z \varphi_i + \chi_i^\dagger Z Z^\dagger \chi_i  + \tilde{\varphi}_i Z Z^\dagger \tilde{\varphi}_i^\dagger - \tilde{\chi}_iZ^\dagger Z \tilde{\chi}_i^\dagger \nn\\ &&\ \  +  {\rm Tr}\,(\phi_i\chi_i^\dagger - \tilde{\chi}_i\tilde{\varphi}_i)(\chi_j\varphi_j^\dagger - \tilde{\varphi}_j^\dagger\tilde{\chi}_j) + {\rm Tr}Z^\dagger Z\nn\ee
We see that the potential includes a harmonic trap for excitations as expected from the form of $C$ given in \eqn{c}

\para
The Lagrangian \eqn{matrixlag} has a number of  symmetries. Rotations in the spatial plane act by $Z\rightarrow e^{i\theta} Z$, with other fields left invariant. The corresponding orbital angular momentum is
\be {\mathbb J}_0= {\rm Tr}\, Z^\dagger Z\nn\ee
There is also a 
 $G\cong S[U(N)\times U(N)] \times U(1)_R$ symmetry, under which $Z$ is a singlet while $\varphi$ transforms as $({\bf N},\bar{\bf N})_1$, $\chi$ as $({\bf N},\bar{\bf N})_{-1}$, 
 $\tilde{\varphi}$ as $(\bar{\bf N},{\bf N})_1$ and $\tilde{\chi}$ as $(\bar{\bf N},{\bf N})_{-1}$.  The $S[U(N)\times U(N)]$ symmetry descends from the $S[U(N_c) \times SU(N_f)]$ gauge and flavour symmetry in $d=2+1$ with $N_c=N_f=N$. This includes the action from the global part of the gauge group, including the overall  $U(1)_N \subset U(N_c)$ whose charge is counted by the operator
 \be {\mathbb N} = \sum_i \varphi_i^\dagger \varphi_i + \chi_i^\dagger\chi_i + {\rm Tr}\,(\tilde{\varphi}_i^\dagger\tilde{\varphi}_i + \tilde{\chi}_i^\dagger\tilde{\chi}_i)\nn\ee
 %
%
%
%
Finally, there are two fermionic supercharges, ${\mathbb Q}$ and ${\mathbb S}$ although, as we describe, only one of them is associated to a symmetry of \eqn{matrixlag}. They descend from $q$ and ${\cal S}$ in $d=2+1$ dimensions respectively and are given by
\be {\mathbb  Q} = \varphi_i^\dagger \chi_i + \tilde{\chi}_i\tilde{\varphi}_i^\dagger\ \ \ {\rm and}\ \ \ {\mathbb S} = \varphi_i^\dagger Z^\dagger \chi_i - \tilde{\chi}_i Z^\dagger \tilde{\varphi}_i^\dagger\nn\ee
It is clear that ${\mathbb S}$ gives rise to the non-linear supersymmetry transformation  that we were looking for, in which the action on the fields depends on the coordinate $Z$ of the vortex. These operators obey
\be [{\mathbb L}_0,{\mathbb Q}] = 0 \ \ \ ,\ \ \ [{\mathbb L}_0,{\mathbb S}] = -{\mathbb S}\nn\ee
Note that the second of these tells us that ${\mathbb S}$ is not a symmetry of the theory. The offending term is the ${\mathbb J}_0$ in the final line of \eqn{matrixlag}; in the absence of this term ${\mathbb S}$ is a symmetry. However, to describe the vortex dynamics the algebra above is required so that it mimics \eqn{lqs}. The anti-commutators of the supercharges give
\be  \{{\mathbb Q},{\mathbb Q}^\dagger\} = {\mathbb N} \ \ \ ,\ \ \ \{ {\mathbb S},{\mathbb S}^\dagger\} = {\mathbb L}_0-{\mathbb J}_0\nn\ee
which should be compared with \eqn{qsnew}. We also have $\{{\mathbb Q},{\mathbb S}^\dagger\} =0$, a relation which requires the use of Gauss' law \eqn{another}.

\para
The symmetries described above coincide with those expected of the low-energy dynamics of vortices. For this reason, we conjecture that it captures some aspects of the dynamics. It would be interesting to understand this connection better and, indeed, to understand the restriction of this type of supersymmetry on the more direct non-linear sigma-model approach to the dynamics of Jackiw-Pi vortices.

\subsection{Open Questions}

This section has discussed a number of properties of Jackiw-Pi vortices. However, their full role in the quantum dynamics of anyons remains far from clear. In the absence of any definitive answers, here we at least try to formulate some definitive questions.

\para
First, a comment. Usually, in relativistic theories, solitons should not be thought of as objects with fixed particle number. Instead, they are coherent states which involve sums over different numbers of particles. They are good approximations to the underlying quantum states when the occupation numbers are large. In the present context of non-relativistic theories, it is not clear that this interpretation continues to hold. The particle number is a super-selection sector and the quantum mechanics describing different numbers of particles do not talk to each other. For this reason we suggest that, in contrast to the usual interpretation (one echoed in \cite{jpi}), the Jackiw-Pi vortices have definite particle number, as opposed to just an average particle number. They are again trustworthy approximations to the underlying quantum states when the occupation numbers are large. 

\para
Further, Jackiw-Pi vortices do not carry any conserved topological quantum number. In this sense, there is nothing to distinguish them from the perturbative states. The following questions are predicated on the assumption that vortices do indeed carry fixed particle number and provide an alternative way of thinking about the perturbative states in certain regimes.

\begin{itemize}
\item What is the significance of the fact that  Jackiw-Pi vortices in the plane first arise at $n=2|k|$ when the operator $\Phi^{2|k|}$ hits the unitarity bound? In particular, what is the explicit connection, if any, between these classical field configurations and the local operator $\Phi^{2|k|}$?
\item What are the solutions to the equations \eqn{bog} describing BPS vortices in a harmonic potential? What is the moduli space? Presumably the vortices are good approximations to certain states of the quantum mechanics. What are these states? What properties do they have?
\item In Section \ref{susyqmsec}, we introduced a new class of supersymmetric gauged linear sigma models, arising from the dimensional reduction of the theory first described in \cite{llm}. These models carry the right symmetries to describe the dynamics of Jackiw-Pi vortices. What aspects of this dynamics do they reliably capture? What are the constraints of supersymmetry on non-linear sigma models of this type? Given the vast array of applications of traditional gauged linear models, are there any further uses of this new class of supersymmetric quantum mechanics?
\end{itemize}

We think these are interesting questions.

\section*{Acknowledgements}

Thanks to Nick Dorey, Ben Gripaios, Siavash Golkar, Kimyeong Lee, Sangmin Lee and Sungjay Lee for useful discussions. DT and CT thank KIAS for their kind hospitality. 
 We are supported by STFC and by the European Research Council under the European Union's Seventh Framework Programme (FP7/2007-2013), ERC grant agreement STG 279943, ``Strongly Coupled Systems".

\appendix

\section{Appendix: Perturbative Analysis}\label{appa}

In this appendix, we present a number of one-loop computations in the non-relativistic Lagrangian \eqn{lag} and related theories. We start by writing down the Feynman rules for these theories. We subsequently describe the requirements for conformal invariance in Section \ref{appa1} and the computation of anomalous dimensions in Section \ref{appa2}.

\para
The Chern-Simons action for the gauge field must be augmented with a gauge fixing term. Combined, they read\footnote{Conventions revisited: the subscripts $\mu,\nu,\rho=0,1,2$ run over both space and time indices, while $a=1,2$ runs over spatial indices only. $i,j=1,\ldots,N_f$ labels the flavour and $\alpha,\beta = 1,\ldots, N_c$ colour. The generators of the gauge group are $T^I_{\alpha\beta}$, with $I,J=1,\ldots,N_c^2$, obeying the Lie algebra $[T^I, T^J]  = i f^{IJK} T^K$.}
\be 
{\cal L}_{\rm gauge} =  -\frac{k}{4\pi} \epsilon^{\mu\nu\rho}\Tr\left\{A_{\mu}\partial_{\nu}A_{\rho} - \frac{2i}{3} A_{\mu}A_{\nu}A_{\rho} \right\}
 + \frac{1}{\xi} \Tr\left\{ \left(\partial_{a} A^{a} \right)^{2} \right\} 
 \nn\ee
In the limit $\xi \rightarrow 0$ the gauge fixing term imposes the Coulomb gauge $\partial_a A^a = 0$. We work in momentum space, with the Fourier transform conventions 
\be
A_{\mu}(t,x) = \int \frac{\rmd^{3}p}{(2\pi)^{3}}\ A_{\mu}(p)\, e^{-i p^{0}t+i \bfp\cdot\bfx}
\nn\ee
%
%
%
The momentum space propagator for the gauge field can be expressed as
\be
D_{\mu\nu}(p^0,\bfp) = \frac{i\xi}{2 (\bfp^{2})^{2} }p_{\mu}p_{\nu} - \frac{2\pi}{k}\epsilon_{\mu\nu a}p^{a}\nn\ee
In what follows, we enforce Coulomb gauge by taking the limit $\xi\rightarrow 0 $. The only non-vanishing components of the propagator   are 
\be
D_{0a} (\bfp) = \frac{2\pi}{k}\,\frac{p^{c}\epsilon_{ca}}{\bfp^2}\nn
\ee
and $D_{a0}=-D_{0a}$. They do not depend on the energy $p^0$, but only on the spatial momenta $\bfp$. 
The resulting momentum space Feynman rules are then
\begin{fmffile}{FR_CS}
\begin{equation}
\label{FR_CS}
\begin{aligned}
\\[5pt]
	\begin{aligned}
	\begin{fmfgraph*}(15,20)
	\fmfleft{i}
	\fmfright{o}
	\fmflabel{$0,I$}{i}
	\fmflabel{$a,J$}{o}
	\fmf{photon}{i,o}
	\end{fmfgraph*}
	\end{aligned}
	\hspace{30pt}&
	\begin{aligned}
	= \delta^{IJ} D_{0a}(p)
	\end{aligned}
\hspace{50pt}
	\begin{aligned}
	\begin{fmfgraph*}(20,20)
	\fmfleft{i1,i2}
	\fmfright{o1,o2}
	\fmf{photon}{i2,v1}
	\fmf{photon,tension=1.2}{v1,v2}
	\fmf{photon}{v1,o2}
	\fmf{phantom}{i1,v2}
	\fmf{phantom}{v2,o1}
	\fmf{phantom,tension=0}{v1,o1}
	\fmflabel{$\mu,I$}{i2}
	\fmflabel{$\nu,J$}{o2}
	\fmflabel{$\rho,K$}{v2}
	\end{fmfgraph*}
	\end{aligned}
	\hspace{-10pt}&
	\begin{aligned}
	=-\frac{ik}{2\pi}\epsilon^{\mu\nu\rho}f^{IJK}
	\end{aligned}
\end{aligned}
\nn\end{equation}
\end{fmffile}
Next we turn to the matter Lagrangian. Here it will prove useful to consider a slightly more general Lagrangian than \eqn{lag}. We retain the general structure of $N_f$ bosons $\phi_i$ and $N_f$ fermions $\psi_i$, each transforming the fundamental representation of $U(N_c)$, now interacting through the Lagrangian
\be
{\cal L}_{\rm matter} &=& i \phi^{\dagger}_{i} D_{t} \phi_{i} + i \psi^{\dagger}_{i} D_{t} \psi_{i}-\frac{1}{2m} |D_{a}\phi_{i}|^{2} - \frac{1}{2m}| D_{a}\psi_{i} |^{2}
+\frac{1}{2m}\psi^{\dagger}_{i} B \psi_i - \frac{\nu}{4} \phi^{\dagger}_{i} \phi_{j} \phi^{\dagger}_{j}\phi_i \nn\\  &&\ \ - \frac{\tilde{\nu}}{4} \left|\phi_{i}\right|^{2} \left|\phi_{j}\right|^{2} 
- \nu_{1} \psi^{\dagger}_{i}\phi_j \phi^{\dagger}_{j}\psi_i -\nu_{2} \phi^{\dagger}_{i} \phi_j \psi^{\dagger}_{j}\psi_i -\tilde{\nu}_{1} \psi^{\dagger}_{i}\phi_i \phi^{\dagger}_{j}\psi_j -\tilde{\nu}_{2} \phi^{\dagger}_{i} \phi_i \psi^{\dagger}_j\psi_j \nn\ee
This includes more general interactions and coupling strengths than the supersymmetric theory \eqn{lag}. We can restrict to the supersymmetric theory by choosing $\tilde{\nu}=\tilde{\nu}_{1}=\tilde{\nu}_{2} = 0$, together with $\nu_{1} = \pi/mk$, $\nu_2 = 2\pi/mk$ and $\nu = 4\pi/mk$.

\para
We can now write down the Feynman rules for the above Lagrangian in momentum space. A solid line denotes a propagating boson while a dashed line represents a propagating fermion. Since the dynamics are non-relativistic, there are no anti-particles and no need to dress the diagrams with arrows.  All fields can only propagate forwards in time which, for us, is left to right. The propagators are
\begin{fmffile}{FR_m1}
\begin{equation}
\begin{aligned}
	&
	\hspace{10pt}
	\begin{aligned}
	\begin{fmfgraph*}(18,20)
	\fmfleft{i}
	\fmfright{o}
	\fmflabel{$i,\alpha$}{i}
	\fmflabel{$j,\beta$}{o}
	\fmf{plain,label.dist=-12,label={$p$}}{i,o}
	\end{fmfgraph*}
	\end{aligned}
	\hspace{25pt}
	\begin{aligned}
	= \delta^{i}_{j}\delta^{\alpha}_{\beta} \, G(p)
	\end{aligned}
	& \hspace{50pt} &
	\hspace{10pt}
	\begin{aligned}
	\begin{fmfgraph*}(18,20)
	\fmfleft{i}
	\fmfright{o}
	\fmflabel{$i,\alpha$}{i}
	\fmflabel{$j,\beta$}{o}
	\fmf{dashes,label.dist=-12,label={$p$}}{i,o}
	\end{fmfgraph*}
	\end{aligned}
	\hspace{25pt}= \delta^{i}_{j}\delta^{\alpha}_{\beta} \, G(p)
\nn\end{aligned}\end{equation}
\end{fmffile}
with 
\be
G(p) = \frac{i}{p^{0}-{\bfp^{2}}/{2m} + i \varepsilon}\label{prop}\ee
\noindent
The coupling of matter to the photon is described by the following four diagrams
\begin{fmffile}{FR_m2}
\begin{equation}
\label{FR_m2}
\begin{aligned}
\\[5pt]	%
	&
	\begin{aligned}
	\begin{fmfgraph*}(25,10)
	\fmfleft{i1,i2}
	\fmfright{o1,o2}
	\fmflabel{$I$}{v1}
	\fmflabel{$\mu$}{v2}
	\fmflabel{$i,\alpha$}{i1}
	\fmflabel{\hspace{-5pt}$j,\beta$}{o1}
	\fmf{plain,label.dist=-11,label={$\bfp$}}{i1,v2}
	\fmf{plain,label.dist=-14,label={$\bfp^{\prime}$}}{v2,o1}
	\fmf{phantom}{i2,v1,o2}
	\fmf{photon,tension=0}{v2,v1}
	\end{fmfgraph*}
	\end{aligned}
	\hspace{20pt}
	\begin{aligned}
	= \delta^{i}_{j} T^{I}_{\alpha\beta} \ \Gamma_{\mu}(\bfp+\bfp^{\prime})
	\end{aligned}
	&\hspace{40pt} &
	\begin{aligned}
	\begin{fmfgraph*}(25,10)
	\fmfleft{i1,i2}
	\fmfright{o1,o2}
	\fmflabel{$I$}{v1}
	\fmflabel{$\mu$}{v2}
	\fmflabel{$i,\alpha$}{i1}
	\fmflabel{\hspace{-5pt}$j,\beta$}{o1}
	\fmf{dashes,label.dist=-11,label={$\bfp$}}{i1,v2}
	\fmf{dashes,label.dist=-14,label={$\bfp^{\prime}$}}{v2,o1}
	\fmf{phantom}{i2,v1,o2}
	\fmf{photon,tension=0}{v2,v1}
	\end{fmfgraph*}
	\end{aligned}
	\hspace{20pt}
	\begin{aligned}
	\\
	=&\ \delta^{i}_{j} T^{I}_{\alpha\beta} \ \Gamma_{\mu}(\bfp+\bfp^{\prime})
	\\
	&
	+\delta^{i}_{j} T^{I}_{\alpha\beta} \ \tilde{\Gamma}_{\mu}(\bfp-\bfp^{\prime})
	\end{aligned}
	\\[40pt]
	&
	\begin{aligned}
	\begin{fmfgraph*}(25,10)
	\fmfleft{i1,i2}
	\fmfright{o1,o2}
	\fmflabel{$a,b$}{v}
	\fmflabel{$i,\alpha$}{i1}
	\fmflabel{$I$}{i2}
	\fmflabel{$J$}{o2}
	\fmflabel{\hspace{-5pt}$j,\beta$}{o1}
	\fmf{plain}{i1,v,o1}
	\fmf{photon,tension=0}{i2,v}
	\fmf{photon,tension=0}{v,o2}
	\end{fmfgraph*}
	\end{aligned}
	\hspace{20pt}
	\begin{aligned}
	= \delta^{i}_{j} \left(T^{(I}T^{J)}\right)_{\alpha\beta}  \, \Gamma_{ab}
	\end{aligned}
	& &
	\begin{aligned}
	\begin{fmfgraph*}(25,10)
	\fmfleft{i1,i2}
	\fmfright{o1,o2}
	\fmflabel{$a,b$}{v}
	\fmflabel{$i,\alpha$}{i1}
	\fmflabel{$I$}{i2}
	\fmflabel{$J$}{o2}
	\fmflabel{\hspace{-5pt}$j,\beta$}{o1}
	\fmf{dashes}{i1,v,o1}
	\fmf{photon,tension=0}{i2,v}
	\fmf{photon,tension=0}{v,o2}
	\end{fmfgraph*}
	\end{aligned}
	\hspace{20pt}
	\begin{aligned}
	\\
	=&\ \delta^{i}_{j} \left(T^{(I}T^{J)}\right)_{\alpha\beta} \, \Gamma_{ab} 
	\\
	&
	+ \delta^{i}_{j} \left(T^{[I}T^{J]}\right)_{\alpha\beta} \, \tilde{\Gamma}_{ab} 
	\end{aligned}
	\\[10pt]
\end{aligned}
\nn\end{equation}
\end{fmffile}
$\! \!\!\!$ with the interaction strengths given by
\be &&\ \ \ \ \ \ \ 
\Gamma_0=\tilde{\Gamma}_0 = i\ \ \ ,\ \ \  
\Gamma_{ab} = -\frac{i}{m}\delta_{ab} \ \ \ ,\ \ \ 
\tilde{\Gamma}_{ab} = \frac{1}{m}\epsilon_{ab}\nn\\ 
&& \Gamma_{a}(\bfp+\bfp^{\prime}) = \frac{i}{2m}(p_{a} + p^{\prime}_{a})\ \ \ ,\ \ \ \tilde{\Gamma}_{a}(\bfp-\bfp^{\prime}) = \frac{1}{2m}\epsilon_{ab}(p^{b}+p^{\prime\,b})
\nn\ee
Finally, the quartic interaction vertices are
\begin{fmffile}{FR_m3}
\begin{equation}
\begin{aligned}
	\\[10pt]
	&&
	\begin{aligned}
	\begin{fmfgraph*}(25,15)
	\fmfleft{i1,i2}
	\fmfright{o1,o2}
	\fmflabel{$i,\alpha$}{i2}
	\fmflabel{$j,\beta$}{i1}
	\fmflabel{$i^{\prime},\alpha^{\prime}$}{o2}
	\fmflabel{$j^{\prime},\beta^{\prime}$}{o1}
	\fmf{plain}{i1,v,o1}
	\fmf{plain}{i2,v,o2}
	\end{fmfgraph*}
	\end{aligned}
	\hspace{20pt}
	\begin{aligned}
	= \Gamma^{ii^{\prime},\alpha\alpha^{\prime}}_{jj^{\prime},\beta\beta^{\prime}}
	\end{aligned}
	&\hspace{70pt} &
	\begin{aligned}
	\begin{fmfgraph*}(25,15)
	\fmfleft{i1,i2}
	\fmfright{o1,o2}
	\fmflabel{$i,\alpha$}{i2}
	\fmflabel{$j,\beta$}{i1}
	\fmflabel{$i^{\prime},\alpha^{\prime}$}{o2}
	\fmflabel{$j^{\prime},\beta^{\prime}$}{o1}
	\fmf{dashes}{i1,v,o1}
	\fmf{plain}{i2,v,o2}
	\end{fmfgraph*}
	\end{aligned}
	\hspace{20pt}
	\begin{aligned}
	= \tilde{\Gamma}^{ii^{\prime},\alpha\alpha^{\prime}}_{jj^{\prime},\beta\beta^{\prime}}
	\end{aligned}
	\\[10pt]
	\nn\end{aligned}\end{equation}
\end{fmffile}
where
\be
i\Gamma^{ii^{\prime},\alpha\alpha^{\prime}}_{jj^{\prime},\beta\beta^{\prime}} &=&   \frac{\nu}{2}\left( \delta^{ii^{\prime}}\delta^{jj^{\prime}}\delta_{\alpha\beta^{\prime}}\delta_{\beta\alpha^{\prime}} +\delta^{ij^{\prime}}\delta^{ji^{\prime}}\delta_{\alpha\alpha^{\prime}}\delta_{\beta\beta^{\prime}} \right)
+\frac{\tilde{\nu}}{2}\left( \delta^{ii^{\prime}}\delta^{jj^{\prime}}\delta_{\alpha\alpha^{\prime}}\delta_{\beta\beta^{\prime}} 
+ \delta^{ij^{\prime}}\delta^{ji^{\prime}}\delta_{\alpha\beta^{\prime}}\delta_{\beta\alpha^{\prime}} \right)
\nn\\[5pt]
i\tilde{\Gamma}^{ii^{\prime},\alpha\alpha^{\prime}}_{jj^{\prime},\beta\beta^{\prime}} &=&
\nu_{1} \delta^{ii^{\prime}}\delta^{jj^{\prime}}\delta_{\alpha\beta^{\prime}}\delta_{\beta\alpha^{\prime}}
+ \nu_{2} \delta^{ij^{\prime}}\delta^{ji^{\prime}}\delta_{\alpha\alpha^{\prime}}\delta_{\beta\beta^{\prime}}
 + \tilde{\nu}_{1} \delta^{ij^{\prime}}\delta^{ji^{\prime}}\delta_{\alpha\beta^{\prime}}\delta_{\beta\alpha^{\prime}}
+ \tilde{\nu}_{2} \delta^{ii^{\prime}}\delta^{jj^{\prime}}\delta_{\alpha\alpha^{\prime}}\delta_{\beta\beta^{\prime}} 
\nn\ee
In the rest of this appendix, we employ these Feynman rules in a number of one-loop computations.  Before we proceed, note that 
 the only objects depending on the loop energy $p^{0}$ are the bosons and fermions running in the loop with propagator \eqn{prop}. The integration over $p^{0}$ may then be carried out by choosing an appropriate contour. As we shall see, all relevant diagrams at one loop have either one or two bosons or fermions running in the loop. The loop energy integral for a single boson or fermion running in the loop is
\be 
\int \frac{\rmd p^{0}}{2\pi}\ G(p) = \frac{1}{2}
\label{intG}\ee
due to a contribution from the pole at $p^{0}=\infty$. For diagrams with two fermions or bosons running in the loop the loop energy integral takes the form
\be
\int \frac{\rmd p^{0}}{2\pi}\ G(p) G(p_{1}+p_{2} - p) = \frac{ -i m}{\bfp^{2}+\bfp_{1}\cdot\bfp_{2} - \bfp\cdot(\bfp_{1}+\bfp_{2}) -i\varepsilon}\,,
\nn\ee
where we have used the non-relativistic on-shell relation $\bfp_{i}^{2}=2mp_{i}^{0}$. In the centre of mass frame $p^{0}=p_{1}^{0}=p_{2}^{0}$ and we can write $\bfp_{1}=-\bfp_{2}$. In this case, we have 
\be
\int \frac{\rmd p^{0}}{2\pi} G(p) G(p_{1}+p_{2} - p) = \frac{ -i m}{\bfp^{2}-\bfp_1^{2} -i\varepsilon}\,.
\label{intGGcm}\ee
In the following section we always choose the centre of mass frame and carry out the loop energy integral using \eqref{intG} and \eqref{intGGcm}.

\subsection{Conformal Fixed Points}\label{appa1}

Our first goal is to determine the values of the coupling constants for which the theory is scale invariant at one-loop. This was first done for the Abelian theory in \cite{bergman} and, for bosons in the the non-Abelian theory, in \cite{bbak}. Here we extend these results to include both bosons and fermions with non-trivial colour and flavour index structures. The upshot of this analysis is that we will find that the supersymmetric theory \eqn{lag} remains conformal at one-loop. However, we will also find other fixed points.

\para
Our strategy is to evaluate the one-loop corrections to the quartic interactions. These will typically exhibit an ultraviolet divergence which, after renormalisation, induces a scale into the theory. We require that all such divergences cancel.

\para
\underline{$\langle\psi\psi\psi^{\dagger}\psi^{\dagger}\rangle$}
\para

We begin the analysis with the $2\rightarrow 2$ scattering of fermions at one-loop. In the absence of any bare four-fermi interaction,  the one-loop correction must be scale invariant for the theory to exhibit conformal symmetry. Let the following diagram denote the one-loop 4-fermion vertex in the centre of mass frame
\begin{fmffile}{4fermion}
\\[10pt]
\begin{equation}
	\begin{aligned}
	\begin{fmfgraph*}(18,13)
	\fmfleft{i1,i2}
	\fmfright{o1,o2}
	\fmflabel{$i,\alpha,\bfp$}{i2}
	\fmflabel{$j,\beta,-\bfp$}{i1}
	\fmflabel{$i^{\prime},\alpha^{\prime},\bfp^{\prime}$}{o2}
	\fmflabel{$j^{\prime},\beta^{\prime},-\bfp^{\prime}$}{o1}
	\fmfv{decoration.shape=circle,decoration.filled=shaded}{v}
	\fmf{dashes}{i1,v,o1}
	\fmf{dashes}{i2,v,o2}
	\end{fmfgraph*}
	\end{aligned}
	\qquad
	=
	\begin{aligned}
	\langle\psi^{\alpha^{\prime}}_{i^{\prime}}(p^{0},\bfp^{\prime}) \psi^{\beta^{\prime}}_{j^{\prime}}(p^{0},-\bfp^{\prime}) \psi^{\dagger}_{i\alpha}(p^{0},\bfp)\psi^{\dagger}_{j\beta}(p^{0},-\bfp)\rangle\,.
	\end{aligned}
\nn\end{equation}
\\[10pt]
\end{fmffile}
We will suppress the labels in the diagrams from now on. The contributing diagrams at one-loop are
\begin{fmffile}{Gamma4f}
\begin{equation}
\label{Gamma4f}
\begin{aligned}
	\begin{aligned}
	\begin{fmfgraph*}(18,14)
	\fmfleft{i1,i2}
	\fmfright{o1,o2}
	\fmfv{decoration.shape=circle,decoration.filled=shaded}{v}
	\fmf{dashes}{i1,v,o1}
	\fmf{dashes}{i2,v,o2}
	\end{fmfgraph*}
	\end{aligned}
	\begin{aligned}
	= 
	\end{aligned}
	\begin{aligned}
	\begin{fmfgraph*}(18,14)
	\fmfleft{i1,i2}
	\fmfright{o1,o2}
	\fmf{dashes}{i1,v1,v3,o1}
	\fmf{dashes}{i2,v2,v4,o2}
	\fmf{photon,tension=.8}{v1,v2}
	\fmf{photon,tension=.8}{v3,v4}
	\end{fmfgraph*}
	\end{aligned}
	\begin{aligned}
	+ 
	\end{aligned}
	\begin{aligned}
	\begin{fmfgraph*}(20,12)
	\fmfleft{i1,i2}
	\fmfright{o1,o2}
	\fmf{dashes}{i1,v1,v2,o1}
	\fmf{dashes}{i2,v,o2}
	\fmf{photon,tension=.1}{v1,v,v2}
	\end{fmfgraph*}
	\end{aligned}
	\begin{aligned}
	+ 
	\end{aligned}
	\begin{aligned}
	\begin{fmfgraph*}(20,12)
	\fmfleft{i1,i2}
	\fmfright{o1,o2}
	\fmf{dashes}{i1,v,o1}
	\fmf{dashes}{i2,v1,v2,o2}
	\fmf{photon,tension=.1}{v1,v,v2}
	\end{fmfgraph*}
	\end{aligned}
	\begin{aligned}
	+ 
	\end{aligned}
	\begin{aligned}
	\begin{fmfgraph*}(20,12)
	\fmfleft{i1,i2}
	\fmfright{o1,o2}
	\fmf{dashes}{i1,v1,v2,o1}
	\fmf{dashes}{i2,v3,o2}
	\fmf{photon,tension=.1}{v1,v,v2}
	\fmf{photon,tension=.2}{v,v3}
	\end{fmfgraph*}
	\end{aligned}
	\begin{aligned}
	+ 
	\end{aligned}
	\begin{aligned}
	\begin{fmfgraph*}(20,12)
	\fmfleft{i1,i2}
	\fmfright{o1,o2}
	\fmf{dashes}{i1,v3,o1}
	\fmf{dashes}{i2,v1,v2,o2}
	\fmf{photon,tension=.1}{v1,v,v2}
	\fmf{photon,tension=.2}{v,v3}
	\end{fmfgraph*}
	\end{aligned}
\end{aligned}
\end{equation}
\end{fmffile}
\noindent
where we have omitted the diagrams where the outgoing external legs are exchanged. Accounting for the anti-commuting nature of the fields, such diagrams are related to the ones above via
\begin{fmffile}{crossed-diagrams}
\\[5pt]
\begin{equation}
	\begin{aligned}
	\begin{fmfgraph*}(18,14)
	\fmfleft{i1,i2}
	\fmfright{o1,o2}
	\fmflabel{$i^{\prime},\alpha^{\prime},\bfp^{\prime}$}{o2}
	\fmflabel{$j^{\prime},\beta^{\prime},-\bfp^{\prime}$}{o1}
	\fmf{dashes,tension=3}{i1,v1,v3}
	\fmf{dashes,tension=3}{i2,v2,v4}
	\fmf{dashes,tansion=0}{v3,o2}
	\fmf{dashes,tension=0}{v4,o1}
	\fmf{phantom,tension=1.5}{v3,o1}
	\fmf{phantom,tension=1.5}{v4,o2}
	\fmf{photon,tension=.8}{v1,v2}
	\fmf{photon,tension=.8}{v3,v4}
	\end{fmfgraph*}
	\end{aligned}
	\begin{aligned}
	\hspace{60pt} = \quad - \quad
	\end{aligned}
	\begin{aligned}
	\begin{fmfgraph*}(18,14)
	\fmflabel{$i^{\prime},\alpha^{\prime},\bfp^{\prime}$}{o1}
	\fmflabel{$j^{\prime},\beta^{\prime},-\bfp^{\prime}$}{o2}
	\fmfleft{i1,i2}
	\fmfright{o1,o2}
	\fmf{dashes}{i1,v1,v3,o1}
	\fmf{dashes}{i2,v2,v4,o2}
	\fmf{photon,tension=.8}{v1,v2}
	\fmf{photon,tension=.8}{v3,v4}
	\end{fmfgraph*}
	\end{aligned}\\[20pt]
\nn\end{equation}
\end{fmffile}
and similarly for the other diagrams. The box diagram in \eqn{Gamma4f} can be expressed as
\begin{fmffile}{Gamma4f1}
\begin{equation}
\label{Gamma4f1}
\begin{aligned}
	\begin{aligned}
	\begin{fmfgraph*}(18,14)
	\fmfleft{i1,i2}
	\fmfright{o1,o2}
	\fmf{dashes}{i1,v1,v3,o1}
	\fmf{dashes}{i2,v2,v4,o2}
	\fmf{photon,tension=.8}{v1,v2}
	\fmf{photon,tension=.8}{v3,v4}
	\end{fmfgraph*}
	\end{aligned}
	\begin{aligned}
	= \delta^{i}_{i^{\prime}}\delta^{j}_{j^{\prime}} T^{I}_{\alpha\gamma} T^{I}_{\beta\kappa} T^{J}_{\gamma \alpha^{\prime}}T^{J}_{\kappa\beta^{\prime}}
	\int \frac{d^2q}{(2\pi)^{2}} \frac{-im}{\bfq^{2}-\bfp^{2}-i\varepsilon} 
	I_{f}^{\text{box}}(\bfp,\bfq,\bfp^{\prime})
	\end{aligned}
\end{aligned}
\nn\end{equation}
\end{fmffile}
where
\be
I_{f}^{\text{box}} &=& \Gamma_{f}^{\mu}(\bfp,\bfq)D_{\mu\nu}(\bfp-\bfq)\Gamma_{f}^{\nu}(-\bfp,-\bfq) \,
		 \Gamma_{f}^{\rho}(\bfq,\bfp^{\prime})D_{\rho\sigma}(\bfq-\bfp^{\prime})\Gamma_{f}^{\sigma}(-\bfq,-\bfp^{\prime})
\nn\\ &=& -\frac{(2\pi)^{2}}{m^{2}q^{2}}\left( 1+2i\frac{\bfq\times\bfp}{(\bfq-\bfp)^{2}} \right)
		 \left(1- 2i\frac{\bfq\times\bfp^{\prime}}{(\bfq-\bfp^{\prime})^{2}} \right)\,,
\nn\ee
and we have introduced the notation $\Gamma_{f}^{\mu}(\bfp,\bfq)=\Gamma^{\mu}(\bfp+\bfq)+\tilde{\Gamma}^{\mu}(\bfp-\bfq)$. The box diagram is therefore logarithmically divergent. We introduce a UV cut-off $\Lambda$, defined as
\be 
\bfq^{2}=2m\Lambda\nn\ee
Note that $\Lambda$ is a cut-off in energy. (This should be borne in mind when comparing to the results of \cite{son} who use a momentum cut-off.) 
The leading contribution from this diagram is then
\begin{fmffile}{Gamma4f1F}
\begin{equation}
\label{Gamma4f1F}
\begin{aligned}
	\begin{aligned}
	\begin{fmfgraph*}(18,14)
	\fmfleft{i1,i2}
	\fmfright{o1,o2}
	\fmf{dashes}{i1,v1,v3,o1}
	\fmf{dashes}{i2,v2,v4,o2}
	\fmf{photon,tension=.8}{v1,v2}
	\fmf{photon,tension=.8}{v3,v4}
	\end{fmfgraph*}
	\end{aligned}
	\begin{aligned}
	= \delta^{i}_{i^{\prime}}\delta^{j}_{j^{\prime}} \delta_{\alpha\alpha^{\prime}}\delta_{\beta\beta^{\prime}} \frac{i\pi}{mk^{2}} \log\frac{\Lambda}{\mu} +\cO(1)\,.
	\end{aligned}
\end{aligned}
\nn\end{equation}
\end{fmffile}
The second and third diagrams in \eqref{Gamma4f} yield identical leading contributions
\begin{fmffile}{Gamma4f2F}
\begin{equation}
\label{Gamma4f2F}
\begin{aligned}
	\begin{aligned}
	\begin{fmfgraph*}(20,12)
	\fmfleft{i1,i2}
	\fmfright{o1,o2}
	\fmf{dashes}{i1,v1,v2,o1}
	\fmf{dashes}{i2,v,o2}
	\fmf{photon,tension=.1}{v1,v,v2}
	\end{fmfgraph*}
	\end{aligned}
	\begin{aligned}
	=
	\end{aligned}
	\begin{aligned}
	\begin{fmfgraph*}(20,12)
	\fmfleft{i1,i2}
	\fmfright{o1,o2}
	\fmf{dashes}{i1,v,o1}
	\fmf{dashes}{i2,v1,v2,o2}
	\fmf{photon,tension=.1}{v1,v,v2}
	\end{fmfgraph*}
	\end{aligned}
	\begin{aligned}
	= \delta^{i}_{i^{\prime}}\delta^{j}_{j^{\prime}} \left( \delta_{\alpha\alpha^{\prime}}\delta_{\beta\beta^{\prime}} + N_{c}\, \delta_{\alpha\beta^{\prime}} \delta_{\beta \alpha^{\prime}} \right) \frac{-i\pi}{4mk^{2}} \log\frac{\Lambda}{\mu} +\cO(1)\,,
	\end{aligned}
\end{aligned}
\nn\end{equation}
\end{fmffile}
while the last two diagrams evaluate to
\begin{fmffile}{Gamma4f3F}
\begin{equation}
\label{Gamma4f3F}
\begin{aligned}
	\begin{aligned}
	\begin{fmfgraph*}(20,12)
	\fmfleft{i1,i2}
	\fmfright{o1,o2}
	\fmf{dashes}{i1,v1,v2,o1}
	\fmf{dashes}{i2,v3,o2}
	\fmf{photon,tension=.1}{v1,v,v2}
	\fmf{photon,tension=.2}{v,v3}
	\end{fmfgraph*}
	\end{aligned}
	\begin{aligned}
	=
	\end{aligned}
	\begin{aligned}
	\begin{fmfgraph*}(20,12)
	\fmfleft{i1,i2}
	\fmfright{o1,o2}
	\fmf{dashes}{i1,v3,o1}
	\fmf{dashes}{i2,v1,v2,o2}
	\fmf{photon,tension=.1}{v1,v,v2}
	\fmf{photon,tension=.2}{v,v3}
	\end{fmfgraph*}
	\end{aligned}
	\begin{aligned}
	= \delta^{i}_{i^{\prime}}\delta^{j}_{j^{\prime}} \left( \delta_{\alpha\alpha^{\prime}}\delta_{\beta\beta^{\prime}} - N_{c}\, \delta_{\alpha\beta^{\prime}} \delta_{\beta\alpha^{\prime}} \right) \frac{-i\pi}{4mk^{2}} \log\frac{\Lambda}{\mu} +\cO(1)\,.
	\end{aligned}
\end{aligned}
\nn\end{equation}
\end{fmffile}
We learn that the scale-dependent contributions to \eqn{Gamma4f} happily cancel each other.

\para
\underline{$\langle\phi\phi\phi^{\dagger}\phi^{\dagger}\rangle$}
\para

The analysis of the $2\rightarrow 2$ scattering of bosons proceeds parallel to the previous case. There are now extra diagrams contributing to this process at one-loop due to the presence of the four-point coupling between bosons, 
\begin{fmffile}{Gamma4}
\begin{equation}
\label{Gamma4}
\begin{aligned}
	\begin{aligned}
	\begin{fmfgraph*}(18,14)
	\fmfleft{i1,i2}
	\fmfright{o1,o2}
	\fmfv{decoration.shape=circle,decoration.filled=shaded}{v}
	\fmf{plain}{i1,v,o1}
	\fmf{plain}{i2,v,o2}
	\end{fmfgraph*}
	\end{aligned}
	\begin{aligned}
	= 
	\end{aligned}
	&
	\begin{aligned}
	\begin{fmfgraph*}(18,14)
	\fmfleft{i1,i2}
	\fmfright{o1,o2}
	\fmf{plain}{i1,v1,v3,o1}
	\fmf{plain}{i2,v2,v4,o2}
	\fmf{photon,tension=.8}{v1,v2}
	\fmf{photon,tension=.8}{v3,v4}
	\end{fmfgraph*}
	\end{aligned}
	\begin{aligned}
	+ 
	\end{aligned}
	\begin{aligned}
	\begin{fmfgraph*}(18,14)
	\fmfleft{i1,i2}
	\fmfright{o1,o2}
	\fmf{plain}{i1,v1,v,o1}
	\fmf{plain}{i2,v2,v,o2}
	\fmf{photon,tension=0}{v1,v2}
	\end{fmfgraph*}
	\end{aligned}
	\begin{aligned}
	+ 
	\end{aligned}
	\begin{aligned}
	\begin{fmfgraph*}(18,14)
	\fmfleft{i1,i2}
	\fmfright{o1,o2}
	\fmf{plain}{i1,v,v1,o1}
	\fmf{plain}{i2,v,v2,o2}
	\fmf{photon,tension=0}{v1,v2}
	\end{fmfgraph*}
	\end{aligned}
	\begin{aligned}
	+ 
	\end{aligned}
	\begin{aligned}
	\begin{fmfgraph*}(20,14)
	\fmfleft{i1,i2}
	\fmfright{o1,o2}
	\fmf{plain,left,tension=.6}{v1,v2,v1}
	\fmf{plain}{i1,v1,i2}
	\fmf{plain}{o1,v2,o2}
	\end{fmfgraph*}
	\end{aligned}
	\\[20pt]
	&
	\begin{aligned}
	+ 
	\end{aligned}
	\begin{aligned}
	\begin{fmfgraph*}(20,12)
	\fmfleft{i1,i2}
	\fmfright{o1,o2}
	\fmf{plain}{i1,v1,v2,o1}
	\fmf{plain}{i2,v,o2}
	\fmf{photon,tension=.1}{v1,v,v2}
	\end{fmfgraph*}
	\end{aligned}
	\begin{aligned}
	+ 
	\end{aligned}
	\begin{aligned}
	\begin{fmfgraph*}(20,12)
	\fmfleft{i1,i2}
	\fmfright{o1,o2}
	\fmf{plain}{i1,v,o1}
	\fmf{plain}{i2,v1,v2,o2}
	\fmf{photon,tension=.1}{v1,v,v2}
	\end{fmfgraph*}
	\end{aligned}
	\begin{aligned}
	+ 
	\end{aligned}
	\begin{aligned}
	\begin{fmfgraph*}(20,12)
	\fmfleft{i1,i2}
	\fmfright{o1,o2}
	\fmf{plain}{i1,v1,v2,o1}
	\fmf{plain}{i2,v3,o2}
	\fmf{photon,tension=.1}{v1,v,v2}
	\fmf{photon,tension=.2}{v,v3}
	\end{fmfgraph*}
	\end{aligned}
	\begin{aligned}
	+ 
	\end{aligned}
	\begin{aligned}
	\begin{fmfgraph*}(20,12)
	\fmfleft{i1,i2}
	\fmfright{o1,o2}
	\fmf{plain}{i1,v3,o1}
	\fmf{plain}{i2,v1,v2,o2}
	\fmf{photon,tension=.1}{v1,v,v2}
	\fmf{photon,tension=.2}{v,v3}
	\end{fmfgraph*}
	\end{aligned}
\end{aligned}
\end{equation}
\end{fmffile}
Recall that the leading contribution to the fermion box diagram arises from the $\tilde{\Gamma}$ terms which are absent here and one can easily check that the first three diagrams are in fact UV finite. Meanwhile,  the last four diagrams yield leading contributions identical to their fermionic analogue, namely
\begin{fmffile}{Gamma41F}
\begin{equation}
\label{Gamma41F}
\begin{aligned}
	\begin{aligned}
	\begin{fmfgraph*}(20,12)
	\fmfleft{i1,i2}
	\fmfright{o1,o2}
	\fmf{plain}{i1,v1,v2,o1}
	\fmf{plain}{i2,v,o2}
	\fmf{photon,tension=.1}{v1,v,v2}
	\end{fmfgraph*}
	\end{aligned}
	\begin{aligned}
	=
	\end{aligned}
	\begin{aligned}
	\begin{fmfgraph*}(20,12)
	\fmfleft{i1,i2}
	\fmfright{o1,o2}
	\fmf{plain}{i1,v,o1}
	\fmf{plain}{i2,v1,v2,o2}
	\fmf{photon,tension=.1}{v1,v,v2}
	\end{fmfgraph*}
	\end{aligned}
	\begin{aligned}
	= \delta^{i}_{i^{\prime}}\delta^{j}_{j^{\prime}} \left( \delta_{\alpha\alpha^{\prime}}\delta_{\beta\beta^{\prime}} + N_{c}\, \delta_{\alpha\beta^{\prime}} \delta_{\beta\alpha^{\prime}} \right) \frac{-i\pi}{4mk^{2}} \log\frac{\Lambda}{\mu} +\cO(1)\,,
	\end{aligned}
\end{aligned}
\nn\end{equation}
\end{fmffile}
\begin{fmffile}{Gamma42F}
\begin{equation}
\label{Gamma42F}
\begin{aligned}
	\begin{aligned}
	\begin{fmfgraph*}(20,12)
	\fmfleft{i1,i2}
	\fmfright{o1,o2}
	\fmf{plain}{i1,v1,v2,o1}
	\fmf{plain}{i2,v3,o2}
	\fmf{photon,tension=.1}{v1,v,v2}
	\fmf{photon,tension=.2}{v,v3}
	\end{fmfgraph*}
	\end{aligned}
	\begin{aligned}
	=
	\end{aligned}
	\begin{aligned}
	\begin{fmfgraph*}(20,12)
	\fmfleft{i1,i2}
	\fmfright{o1,o2}
	\fmf{plain}{i1,v3,o1}
	\fmf{plain}{i2,v1,v2,o2}
	\fmf{photon,tension=.1}{v1,v,v2}
	\fmf{photon,tension=.2}{v,v3}
	\end{fmfgraph*}
	\end{aligned}
	\begin{aligned}
	= \delta^{i}_{i^{\prime}}\delta^{j}_{j^{\prime}} \left( \delta_{\alpha\alpha^{\prime}}\delta_{\beta\beta^{\prime}} - N_{c}\, \delta_{\alpha\beta^{\prime}} \delta_{\beta\alpha^{\prime}} \right) \frac{-i\pi}{4mk^{2}} \log\frac{\Lambda}{\mu} +\cO(1)\,.
	\end{aligned}
\end{aligned}
\nn\end{equation}
\end{fmffile}
The only diagram left to compute is the diagram with no internal gauge bosons which can easily be seen to yield
\begin{fmffile}{Gamma43F}
\begin{equation}
\label{Gamma43F}
\begin{aligned}
	\begin{aligned}
	\begin{fmfgraph*}(20,14)
	\fmfleft{i1,i2}
	\fmfright{o1,o2}
	\fmf{plain,left,tension=.6}{v1,v2,v1}
	\fmf{plain}{i1,v1,i2}
	\fmf{plain}{o1,v2,o2}
	\end{fmfgraph*}
	\\[20pt]
	\end{aligned}
	\begin{aligned}
	=&\ \bigg[ (\nu^{2}+\tilde{\nu}^{2})\left(\delta^{i}_{i^{\prime}}\delta^{j}_{j^{\prime}}\delta_{\alpha\alpha^{\prime}}\delta_{\beta\beta^{\prime}} 
	+ \delta^{i}_{j^{\prime}}\delta^{j}_{i^{\prime}}\delta_{\alpha\beta^{\prime}}\delta_{\beta\alpha^{\prime}} \right)
	\\
	&\quad + 2\nu\tilde{\nu}\left(\delta^i_{i^\prime}\delta^j_{j^\prime}\delta_{\alpha\beta^{\prime}}\delta_{\beta\alpha^{\prime}} 
	+ \delta^i_{j^\prime}\delta^{j}_{i^{\prime}}\delta_{\alpha\alpha^{\prime}}\delta_{\beta\beta^{\prime}}\right)
	\bigg] \frac{im}{16\pi}\log\frac{\Lambda}{\mu} + \cO(1)\,.
	\end{aligned}
\end{aligned}
\nn\end{equation}
\end{fmffile}
The index structure in the second line cannot be canceled by any other diagram. Scale invariance then forces either $\nu$ or $\tilde{\nu}$ to vanish. Setting $\tilde{\nu}=0$, the leading one-loop contribution \eqref{Gamma4} reads
\begin{fmffile}{Gamma4Final}
\begin{equation}
\label{Gamma4Final}
\begin{aligned}
	\begin{aligned}
	\begin{fmfgraph*}(18,14)
	\fmfleft{i1,i2}
	\fmfright{o1,o2}
	\fmfv{decoration.shape=circle,decoration.filled=shaded}{v}
	\fmf{plain}{i1,v,o1}
	\fmf{plain}{i2,v,o2}
	\end{fmfgraph*}
	\end{aligned}
	\begin{aligned}
	= \left(\delta^{ii^{\prime}}\delta^{jj^{\prime}}\delta_{\alpha\alpha^{\prime}}\delta_{\beta\beta^{\prime}} 
	+ \delta^{ij^{\prime}}\delta^{ji^{\prime}}\delta_{\alpha\beta^{\prime}}\delta_{\beta\alpha^{\prime}} \right)
	\left(\frac{im\nu^{2}}{16\pi} - \frac{i\pi}{mk^{2}}\right)\log\frac{\Lambda}{\mu} +\cO(1)
	\end{aligned}
\end{aligned}
\nn\end{equation}
\end{fmffile}
where we have included the contribution from the triangle diagrams with the outgoing external legs exchanged. For the theory to exhibit conformal symmetry at one-loop level we must set
\begin{equation}
\label{nu-kappa}
\nu = \pm \frac{4\pi}{mk}.
\nn\end{equation}
This is in agreement with the result of \cite{bergman} and \cite{bbak}.

\para
\underline{$\langle\phi\psi\phi^{\dagger}\psi^{\dagger}\rangle$}
\para

We end our study of conformal invariance by looking $2\rightarrow 2$ scattering between particles whose spin differs by $1/2$. 
The contributing diagrams at one-loop are
\begin{fmffile}{tGamma4}
\begin{equation}
\label{tGamma4}
\begin{aligned}
	\begin{aligned}
	\begin{fmfgraph*}(18,14)
	\fmfleft{i1,i2}
	\fmfright{o1,o2}
	\fmfv{decoration.shape=circle,decoration.filled=shaded}{v}
	\fmf{dashes}{i1,v,o1}
	\fmf{plain}{i2,v,o2}
	\end{fmfgraph*}
	\end{aligned}
	\begin{aligned}
	= 
	\end{aligned}
	&
	\begin{aligned}
	\begin{fmfgraph*}(18,14)
	\fmfleft{i1,i2}
	\fmfright{o1,o2}
	\fmf{dashes}{i1,v1,v3,o1}
	\fmf{plain}{i2,v2,v4,o2}
	\fmf{photon,tension=.8}{v1,v2}
	\fmf{photon,tension=.8}{v3,v4}
	\end{fmfgraph*}
	\end{aligned}
	\begin{aligned}
	+ 
	\end{aligned}
	\begin{aligned}
	\begin{fmfgraph*}(18,14)
	\fmfleft{i1,i2}
	\fmfright{o1,o2}
	\fmf{dashes}{i1,v1,v,o1}
	\fmf{plain}{i2,v2,v,o2}
	\fmf{photon,tension=0}{v1,v2}
	\end{fmfgraph*}
	\end{aligned}
	\begin{aligned}
	+ 
	\end{aligned}
	\begin{aligned}
	\begin{fmfgraph*}(18,14)
	\fmfleft{i1,i2}
	\fmfright{o1,o2}
	\fmf{dashes}{i1,v,v1,o1}
	\fmf{plain}{i2,v,v2,o2}
	\fmf{photon,tension=0}{v1,v2}
	\end{fmfgraph*}
	\end{aligned}
	\begin{aligned}
	+ 
	\end{aligned}
	\begin{aligned}
	\begin{fmfgraph*}(20,14)
	\fmfleft{i1,i2}
	\fmfright{o1,o2}
	\fmf{plain,left,tension=.6}{v1,v2}
	\fmf{dashes,left,tension=.6}{v2,v1}
	\fmf{plain}{i2,v1}
	\fmf{plain}{v2,o2}
	\fmf{dashes}{i1,v1}
	\fmf{dashes}{v2,o1}
	\end{fmfgraph*}
	\end{aligned}
	\\[20pt]
	&
	\begin{aligned}
	+ 
	\end{aligned}
	\begin{aligned}
	\begin{fmfgraph*}(20,12)
	\fmfleft{i1,i2}
	\fmfright{o1,o2}
	\fmf{dashes}{i1,v1,v2,o1}
	\fmf{plain}{i2,v,o2}
	\fmf{photon,tension=.1}{v1,v,v2}
	\end{fmfgraph*}
	\end{aligned}
	\begin{aligned}
	+ 
	\end{aligned}
	\begin{aligned}
	\begin{fmfgraph*}(20,12)
	\fmfleft{i1,i2}
	\fmfright{o1,o2}
	\fmf{dashes}{i1,v,o1}
	\fmf{plain}{i2,v1,v2,o2}
	\fmf{photon,tension=.1}{v1,v,v2}
	\end{fmfgraph*}
	\end{aligned}
	\begin{aligned}
	+ 
	\end{aligned}
	\begin{aligned}
	\begin{fmfgraph*}(20,12)
	\fmfleft{i1,i2}
	\fmfright{o1,o2}
	\fmf{dashes}{i1,v1,v2,o1}
	\fmf{plain}{i2,v3,o2}
	\fmf{photon,tension=.1}{v1,v,v2}
	\fmf{photon,tension=.2}{v,v3}
	\end{fmfgraph*}
	\end{aligned}
	\begin{aligned}
	+ 
	\end{aligned}
	\begin{aligned}
	\begin{fmfgraph*}(20,12)
	\fmfleft{i1,i2}
	\fmfright{o1,o2}
	\fmf{dashes}{i1,v3,o1}
	\fmf{plain}{i2,v1,v2,o2}
	\fmf{photon,tension=.1}{v1,v,v2}
	\fmf{photon,tension=.2}{v,v3}
	\end{fmfgraph*}
	\end{aligned}
\end{aligned}
\nn\end{equation}
\end{fmffile}
The computation of each of these diagrams is identical to those described above. For this reason, we present only the final result:
\begin{fmffile}{tGamma4F}
\begin{equation}
\label{tGamma4F}
\begin{aligned}
	\begin{aligned}
	\begin{fmfgraph*}(18,14)
	\fmfleft{i1,i2}
	\fmfright{o1,o2}
	\fmfv{decoration.shape=circle,decoration.filled=shaded}{v}
	\fmf{dashes}{i1,v,o1}
	\fmf{plain}{i2,v,o2}
	\end{fmfgraph*}
	\\[95pt]
	\end{aligned}
	\begin{aligned}
	= &\ \Bigg\{
	\delta^i_{i^\prime}\delta^j_{j^\prime}\delta_{\alpha\alpha^{\prime}}\delta_{\beta\beta^{\prime}}
		\left( \nu_{1}^{2}+\nu_{2}^{2}+\tilde{\nu}_{1}^{2}+\tilde{\nu}_{2}^{2} -\frac{2\pi\nu_{1}}{mk}
		- \frac{3\pi^{2}}{m^{2}k^{2} }\right)
		\\
		&\quad +2\delta^i_{i^\prime}\delta^j_{j^\prime}\delta_{\alpha\beta^{\prime}}\delta_{\beta\alpha^{\prime}}
		\left( \nu_{1}\tilde{\nu}_{2}+\nu_{2}\tilde{\nu}_{1}-\frac{\pi}{mk}\tilde{\nu}_{2} \right)
		\\[5pt]
		&\quad +2\delta^i_{j^\prime}\delta^{j}_{i^{\prime}}\delta_{\alpha\alpha^{\prime}}\delta_{\beta\beta^{\prime}}
		\left( \nu_{1}\tilde{\nu}_{1}+\nu_{2}\tilde{\nu}_{2}-\frac{\pi}{mk} \tilde{\nu}_{1} \right)
		\\
		&\quad +2\delta^i_{j^\prime}\delta^{j}_{i^{\prime}}\delta_{\alpha\beta^{\prime}}\delta_{\beta\alpha^{\prime}}
		\left( \nu_{1}\nu_{2}+\tilde{\nu}_{1}\tilde{\nu}_{2}-\frac{\pi}{mk} \nu_{2} \right)
	\Bigg\} \frac{im}{4\pi} \log\frac{\Lambda}{\mu}+\cO(1)\,,
	\end{aligned}
\end{aligned}
\nn\end{equation}
\end{fmffile}
For scale invariance, each of these terms must vanish at a fixed point. This results in the following restrictions on the various coupling constants 
\begin{equation}
\begin{aligned}
\nu_{1}^{2}+\nu_{2}^{2}+\tilde{\nu}_{1}^{2}+\tilde{\nu}_{2}^{2} -\frac{2\pi\nu_{1}}{mk} &= \frac{3\pi^{2}}{m^{2}k^{2} }\nn\\ 
	\nu_{1}\tilde{\nu}_{2}+\nu_{2}\tilde{\nu}_{1}-\frac{\pi}{mk}\tilde{\nu}_{2}=
	\nu_{1}\tilde{\nu}_{1}+\nu_{2}\tilde{\nu}_{2}-\frac{\pi}{mk} \tilde{\nu}_{1} & =
	\nu_{1}\nu_{2}+\tilde{\nu}_{1}\tilde{\nu}_{2}-\frac{\pi}{mk} \nu_2=0
\nn\end{aligned}\end{equation}
These should then be supplemented by the condition from the bosonic sector, 
\be \nu\tilde{\nu}=\nu^{2}+\tilde{\nu}^{2} - \frac{16\pi^{2}}{m^{2}k^{2}} &=\ 0 \nn\ee
We note that the coupling constants of the supersymmetric theory \eqn{lag} solve these equations.

\subsection{Anomalous Dimensions}\label{appa2}

We now turn to the computation of the one-loop anomalous dimension of operators in the supersymmetric theory \eqn{lag}. For the purely bosonic or purely fermionic sector, this computation was first described by Nishida and Son in \cite{son}. In particular, we show that the one-loop anomalous dimensions of chiral primary operators ${\cal O} = (\Phi^\dagger)^n$ and anti-chiral primary operators $\tilde{\cal O} = \Psi^\dagger \partial_z\Psi^\dagger \ldots \partial_z^{n-1} \Psi^\dagger$  saturate the bounds \eqn{cp} and \eqn{acp} respectively.

\para
To compute the anomalous dimension, we follow \cite{son} and  introduce a source for the given operator ${\cal O}$. This is depicted by a circle $\circ$ in the Feynman diagram. In what follows, we will compute the anomalous dimension of operators involving $\phi$ and $\psi$ instead of $\phi^\dagger$ and $\psi^\dagger$. Correspondingly, the sources are really sinks, into which the $n$ particles described by ${\cal O}$ plunge. These diagrams suffer logarithmic divergences in perturbation theory of the form $\alpha \log(\Lambda/\mu)$ where $\Lambda$ is the energy cut-off.  The anomalous dimension of the operator ${\cal O}$ is then given by
\be\gamma = +2\mu\frac{d}{d\mu}\left(\alpha \log\frac{\Lambda}{\mu}\right) = - 2\alpha\nn\ee
For a single-particle operator, such as $\phi$ or $\psi$, there is no renormalisation. Each has dimension $\Delta=1$. An anomalous dimension first arises with $n=2$ particles. 

\subsubsection*{Bosons}

We start by considering the two particle bosonic operator $\phi_i^\alpha\phi_j^\beta$, replete with its flavour indices $i,j=1,\ldots,N_f$ and colour indices $\alpha,\beta=1,\ldots,N_c$. At tree-level, this has dimension $2$. The operator is represented by the following diagram
\begin{fmffile}{phi2bi}
\begin{equation}
\begin{aligned}
	\begin{aligned}
	\begin{fmfgraph*}(15,10)
	\fmfleft{i1,i2}
	\fmfright{o}
	\fmfv{decoration.shape=circle,decoration.filled=empty,decoration.size=9}{v}
	\fmf{phantom,tension=2}{v,o}
	\fmf{plain}{i1,v,i2}
	\end{fmfgraph*}
	\end{aligned}
	\begin{aligned}
	\sim\quad \left(\delta^i_{i^\prime}\delta^j_{j^\prime} \delta_{\alpha\alpha^{\prime}}\delta_{\beta\beta^{\prime}}+\delta^i_{j^\prime}\delta^{j}_{i^{\prime}} \delta_{\alpha\beta^{\prime}}\delta_{\beta\alpha^{\prime}}\right)\,,
	\\
	\end{aligned}
\end{aligned}
\label{bosontree}\end{equation}
\end{fmffile}
Here the $(i',\alpha')$ and $(j',\beta')$ indices refer to the external legs, while the $(i,\alpha)$ and $(j,\beta)$ indices refer to the source itself. The delta-functions are telling us that, quite obviously, these indices must agree. 
Including the one-loop corrections, the source becomes
\begin{fmffile}{phi2}
\begin{equation}
\begin{aligned}
	\begin{aligned}
	\begin{fmfgraph*}(15,10)
	\fmfleft{i1,i2}
	\fmfright{o}
	\fmfv{decoration.shape=circle,decoration.filled=shaded,decoration.size=12}{v}
	\fmf{phantom,tension=2}{v,o}
	\fmf{plain}{i1,v,i2}
	\end{fmfgraph*}
	\end{aligned}
	\begin{aligned}
	= \ \\[4pt]
	\end{aligned}
	\begin{aligned}
	\begin{fmfgraph*}(15,10)
	\fmfleft{i1,i2}
	\fmfright{o}
	\fmfv{decoration.shape=circle,decoration.filled=empty,decoration.size=9}{v}
	\fmf{phantom,tension=2}{v,o}
	\fmf{plain}{i1,v,i2}
	\end{fmfgraph*}
	\end{aligned}
	\begin{aligned}
	+ \ \\[4pt]
	\end{aligned}
	\begin{aligned}
	\begin{fmfgraph*}(15,10)
	\fmfleft{i1,i2}
	\fmfright{o}
	\fmfv{decoration.shape=circle,decoration.filled=empty,decoration.size=9}{v}
	\fmf{phantom,tension=2}{v,o}
	\fmf{plain}{i1,v1,v,v2,i2}
	\fmf{photon,tension=0}{v1,v2}
	\end{fmfgraph*}
	\end{aligned}
	\begin{aligned}
	+ \ \\[4pt]
	\end{aligned}
	\begin{aligned}
	\begin{fmfgraph*}(15,10)
	\fmfleft{i1,i2}
	\fmfright{o}
	\fmfv{decoration.shape=circle,decoration.filled=empty,decoration.size=9}{v}
	\fmf{phantom,tension=2}{v,o}
	\fmf{plain}{i1,v1,i2}
	\fmf{plain,left,tension=.6}{v1,v,v1}
	\end{fmfgraph*}
	\end{aligned}
\end{aligned}
\nn\end{equation}
\end{fmffile}
The first one-loop correction is finite. It does not contribute to the anomalous dimension. The final diagram suffers a logarithmic divergence. This is what we're looking for. We denote the momentum on the external legs as $\bfp_1$ and $\bfp_2$. Using \eqn{intGGcm}, this diagram is given by
\begin{fmffile}{dphi2}
\begin{equation}
\label{dphi2}
\begin{aligned}
	\begin{aligned}
	\begin{fmfgraph*}(15,10)
	\fmfleft{i1,i2}
	\fmfright{o}
	\fmfv{decoration.shape=circle,decoration.filled=empty,decoration.size=9}{v}
	\fmf{phantom,tension=2}{v,o}
	\fmf{plain}{i1,v1,i2}
	\fmf{plain,left,tension=.6}{v1,v,v1}
	\end{fmfgraph*}
	\\[30pt]
	\end{aligned}
	\begin{aligned}
	&= \left(\delta^i_{i^\prime}\delta^j_{j^\prime} \delta_{\alpha\beta^{\prime}}\delta_{\beta\alpha^{\prime}}+\delta^i_{j^\prime}\delta^{j}_{i^{\prime}} \delta_{\alpha\alpha^{\prime}}\delta_{\beta\beta^{\prime}}\right)
	\int \frac{d^{2}q}{(2\pi)^{2}}\, \frac{-2\pi/k}{\bfq^{2}+\bfp_{1}\cdot\bfp_{2} - \bfq\cdot(\bfp_{1}+\bfp_{2}) - i\varepsilon}
	\\
	&= \left(\delta^i_{i^\prime}\delta^j_{j^\prime} \delta_{\alpha\beta^{\prime}}\delta_{\beta\alpha^{\prime}}+\delta^i_{j^\prime}\delta^{j}_{i^{\prime}} \delta_{\alpha\alpha^{\prime}}\delta_{\beta\beta^{\prime}}\right)
	\left(-\frac{1}{2k}\right)\log\frac{\Lambda}{\mu}+ \cO(1)\,.
	\end{aligned}
\end{aligned}
\nn\end{equation}
\end{fmffile}
The index structure of this one-loop diagram contribution differs from the tree level diagram \eqn{bosontree}. This reflects the fact that the operator $\phi_i^\alpha\phi_j^\beta$ sits in a reducible representation of the $U(N_c)\times SU(N_f)$ gauge and flavour groups. We should instead decompose it as
\be {\cal O}_{\rm sym} = \phi^{(i}_{(\alpha}\phi^{j)}_{\beta)}\ \ \ ,\ \ \  {\cal O}_{\rm anti-sym}=\phi^{[i}_{[\alpha}\phi^{j]}_{\beta]} \nn\ee
with the anomalous dimension $\gamma$ given by
\be
\gamma_{\rm sym} =  \frac{1}{k}\ \ \ ,\ \ \ \gamma_{\rm anti-sym}=- \frac{1}{k}\nn \ee
The generalisation to multiple bosons is straightforward since the relevant diagrams are simply pairwise generalisations of the two particle story,
\begin{fmffile}{phin}
\begin{equation}
\begin{aligned}
	\begin{aligned}
	\begin{fmfgraph*}(18,12)
	\fmfleft{i3,i1,i2}
	\fmfright{o2,o1}
	\fmfv{decoration.shape=circle,decoration.filled=shaded,decoration.size=12}{v}
	\fmf{phantom,tension=2}{v,o1}
	\fmf{plain}{i2,v}
	\fmf{phantom}{i3,v3,o2}
	\fmf{plain}{i1,v1,v,v2,v3}
	\fmf{dots,right,tension=0}{v1,v2}
	\end{fmfgraph*}
	\end{aligned}
	\begin{aligned}
	= \ \\[4pt]
	\end{aligned}
	\begin{aligned}
	\begin{fmfgraph*}(18,12)
	\fmfleft{i3,i1,i2}
	\fmfright{o2,o1}
	\fmfv{decoration.shape=circle,decoration.filled=empty,decoration.size=9}{v}
	\fmf{phantom,tension=2}{v,o1}
	\fmf{plain}{i2,v}
	\fmf{phantom}{i3,v3,o2}
	\fmf{plain}{i1,v1,v,v2,v3}
	\fmf{dots,right,tension=0}{v1,v2}
	\end{fmfgraph*}
	\end{aligned}
	\begin{aligned}
	+ \sum_{\text{pairs}} \ \\[4pt]
	\end{aligned}
	\begin{aligned}
	\begin{fmfgraph*}(18,12)
	\fmfleft{i3,i1,i2}
	\fmfright{o2,o1}
	\fmfv{decoration.shape=circle,decoration.filled=empty,decoration.size=9}{v}
	\fmf{phantom,tension=2}{v,o1}
	\fmf{phantom}{i3,v3,o2}
	\fmf{plain}{i2,v1,i1}
	\fmf{plain,left}{v1,v}
	\fmf{plain}{v,v4,v1}
	\fmf{plain}{v,v2,v3}
	\fmf{dots,right,tension=0}{v4,v2}
	\end{fmfgraph*}
	\end{aligned}
\end{aligned}
\nn\end{equation}
\end{fmffile}
where we have excluded finite loop diagrams from photon exchange. 
In particular, for $N_f=1$ flavours of boson in the $U(1)$ theory, we have \cite{son}
\be
\gamma_{\phi^n} =  {n \choose 2} \gamma_{\phi^{2}} = \frac{n(n-1)}{2k}\ \ \ \implies\ \ \ \Delta_{\phi^n} = n + \frac{n(n-1)}{2k}\nn\ee
As described in the main text, this coincides with the expected result \eqn{cpdim} based on the conformal algebra. This ensures that  the result is one-loop exact.

\subsubsection*{Fermions}

We now turn to fermions. For $N_c>1$ or $N_f>1$ we can build anti-chiral operators of the form $\psi^{I}_{i}\psi^{J}_{j}$ without resorting to derivatives.  This is renormalised at one-loop by 
\begin{fmffile}{psi2}
\begin{equation}
\begin{aligned}
	\begin{aligned}
	\begin{fmfgraph*}(15,10)
	\fmfleft{i1,i2}
	\fmfright{o}
	\fmfv{decoration.shape=circle,decoration.filled=shaded,decoration.size=12}{v}
	\fmf{phantom,tension=2}{v,o}
	\fmf{dashes}{i1,v,i2}
	\end{fmfgraph*}
	\end{aligned}
	\begin{aligned}
	= \ \\[4pt]
	\end{aligned}
	\begin{aligned}
	\begin{fmfgraph*}(15,10)
	\fmfleft{i1,i2}
	\fmfright{o}
	\fmfv{decoration.shape=circle,decoration.filled=empty,decoration.size=9}{v}
	\fmf{phantom,tension=2}{v,o}
	\fmf{dashes}{i1,v,i2}
	\end{fmfgraph*}
	\end{aligned}
	\begin{aligned}
	+ \ \\[4pt]
	\end{aligned}
	\begin{aligned}
	\begin{fmfgraph*}(15,10)
	\fmfleft{i1,i2}
	\fmfright{o}
	\fmfv{decoration.shape=circle,decoration.filled=empty,decoration.size=9}{v}
	\fmf{phantom,tension=2}{v,o}
	\fmf{dashes}{i1,v1,v,v2,i2}
	\fmf{photon,tension=0}{v1,v2}
	\end{fmfgraph*}
	\end{aligned}
\end{aligned}
\nn\end{equation}
\end{fmffile}
The loop diagram can be readily computed to be
\begin{fmffile}{dpsi2}
\begin{equation}
\begin{aligned}
	\begin{aligned}
	\begin{fmfgraph*}(15,10)
	\fmfleft{i1,i2}
	\fmfright{o}
	\fmfv{decoration.shape=circle,decoration.filled=empty,decoration.size=9}{v}
	\fmf{phantom,tension=2}{v,o}
	\fmf{dashes}{i1,v1,v,v2,i2}
	\fmf{photon,tension=0}{v1,v2}
	\end{fmfgraph*}
	\end{aligned}
	\begin{aligned}
	= \left(\delta^{ii^{\prime}}\delta^{jj^{\prime}}\delta_{\alpha\beta^{\prime}}\delta_{\beta\alpha^{\prime}}
	-\delta^{ij^{\prime}}\delta^{ji^{\prime}}\delta_{\alpha\alpha^{\prime}}\delta_{\beta\beta^{\prime}}\right)
	\frac{1}{2k}\log\frac{\Lambda}{\mu}+ \cO(1)
	\end{aligned}
\end{aligned}
\nn\end{equation}
\end{fmffile}
This again carries a different index structure from the tree-level diagram. Breaking the operator into irreducible representations of 
 $U(N_{c})\times SU(N_{f})$, we have
\be
\tilde{\cal O}_{\rm flavour-sym} = \psi^{(i}_{[\alpha}\psi^{j)}_{\beta]}\ \ \ ,\ \ \ \tilde{\cal O}_{\rm colour-sym} = \psi^{[i}_{(\alpha}\psi^{j]}_{\beta)}\nn\ee
which have anomalous dimensions
\be \gamma_{\rm flavour-sym} = + \frac{1}{k}\ \ \ ,\ \ \ \gamma_{\rm colour-sym} = - \frac{1}{k}\nn\ee
For multiple fermions, the relevant diagrams again involve a pairwise photon exchange.

\para
In the main text, our primary focus was on the case of  $N_c=N_f=1$. Here, operators involving more than one fermion necessarily involve derivatives.  We  denote a derivative insertion by a cross $\times$ in the diagram. The simplest two-particle operator constructed from fermions involves a single derivative:  $\psi\partial_{a}\psi$. Up to one-loop, the relevant diagrams are
\begin{fmffile}{psidpsi}
\begin{equation}
\begin{aligned}
	\begin{aligned}
	\begin{fmfgraph*}(15,10)
	\fmfleft{i1,i2}
	\fmfright{o}
	\fmfv{decoration.shape=circle,decoration.filled=shaded,decoration.size=12}{v}
	\fmf{phantom,tension=2}{v,o}
	\fmf{dashes}{i1,v1,v,i2}
	\fmfv{decoration.shape=cross,decoration.size=8,decoration.angle=30}{v1}
	\end{fmfgraph*}
	\end{aligned}
	\begin{aligned}
	= \ \\[4pt]
	\end{aligned}
	\begin{aligned}
	\begin{fmfgraph*}(15,10)
	\fmfleft{i1,i2}
	\fmfright{o}
	\fmfv{decoration.shape=circle,decoration.filled=empty,decoration.size=9}{v}
	\fmf{phantom,tension=2}{v,o}
	\fmf{dashes}{i1,v1,v,i2}
	\fmfv{decoration.shape=cross,decoration.size=8,decoration.angle=30}{v1}
	\end{fmfgraph*}
	\end{aligned}
	\begin{aligned}
	+ \ \\[4pt]
	\end{aligned}
	\begin{aligned}
	\begin{fmfgraph*}(15,10)
	\fmfleft{i1,i2}
	\fmfright{o}
	\fmfv{decoration.shape=circle,decoration.filled=empty,decoration.size=9}{v}
	\fmf{phantom,tension=2}{v,o}
	\fmf{dashes}{i1,v1,v3,v,v2,i2}
	\fmfv{decoration.shape=cross,decoration.size=8,decoration.angle=30}{v3}
	\fmf{photon,tension=0}{v1,v2}
	\end{fmfgraph*}
	\end{aligned}
\end{aligned}
\nn\end{equation}
\end{fmffile}
To compute the loop contribution, it is useful to define ${\bf P}^{\pm}=\bfp_{1}\pm\bfp_{2}$. The tree level diagram may then be expressed as
\begin{fmffile}{psidpsitree}
\begin{equation}
\begin{aligned}
	\begin{aligned}
	\begin{fmfgraph*}(15,10)
	\fmfleft{i1,i2}
	\fmfright{o}
	\fmfv{decoration.shape=circle,decoration.filled=empty,decoration.size=9}{v}
	\fmf{phantom,tension=2}{v,o}
	\fmf{dashes}{i1,v1,v,i2}
	\fmfv{decoration.shape=cross,decoration.size=8,decoration.angle=30}{v1}
	\end{fmfgraph*}
	\end{aligned}
	\begin{aligned}
	= -iP^{-}_{a}
	\\[4pt]
	\end{aligned}
\end{aligned}
\nn
\end{equation}
\end{fmffile}
The loop diagram is given by
\begin{fmffile}{psidpsiloop1}
\begin{equation}
\begin{aligned}
	\begin{aligned}
	\begin{fmfgraph*}(15,10)
	\fmfleft{i1,i2}
	\fmfright{o}
	\fmfv{decoration.shape=circle,decoration.filled=empty,decoration.size=9}{v}
	\fmf{phantom,tension=2}{v,o}
	\fmf{dashes}{i1,v1,v3,v,v2,i2}
	\fmfv{decoration.shape=cross,decoration.size=8,decoration.angle=30}{v3}
	\fmf{photon,tension=0}{v1,v2}
	\end{fmfgraph*}
	\end{aligned}
	\begin{aligned}
	= \frac{2i\pi}{k}\int\frac{\rmd^{2}q}{(2\pi)^{2}}\frac{{\bf P}^{+}_{a}-2\bfq_{a}}{\bfq^{2}-(\frac{{\bf P}^{-}}{2})^{2}}
	\left(1+\frac{i\bfq\times{\bf P}^{-}}{(\bfq-\frac{{\bf P}^{-}}{2})^{2}}\right)
	\\[4pt]
	\end{aligned}
\end{aligned}
\nn
\end{equation}
\end{fmffile}
The leading contribution from this diagram takes the form
\begin{fmffile}{psidpsiloop2}
\begin{equation}
\begin{aligned}
	\begin{aligned}
	\begin{fmfgraph*}(15,10)
	\fmfleft{i1,i2}
	\fmfright{o}
	\fmfv{decoration.shape=circle,decoration.filled=empty,decoration.size=9}{v}
	\fmf{phantom,tension=2}{v,o}
	\fmf{dashes}{i1,v1,v3,v,v2,i2}
	\fmfv{decoration.shape=cross,decoration.size=8,decoration.angle=30}{v3}
	\fmf{photon,tension=0}{v1,v2}
	\end{fmfgraph*}
	\end{aligned}
	\begin{aligned}
	= \frac{1}{2k}\epsilon_{ab}({\bf P}^-)^b\log\frac{\Lambda}{\mu}+\cO(1)
	\\[4pt]
	\end{aligned}
\end{aligned}
\nn
\end{equation}
\end{fmffile}
The conformal operators, with well-defined anomalous dimensions, have either holomorphic or anti-holomorphic derivative insertions,
\begin{equation}
\gamma_{\psi\partial\psi} = \frac{1}{k} \ \ \ ,\ \ \ 
\gamma_{\psi\bar{\partial}\psi} = -\frac{1}{k}
\nn
\end{equation}
The operator $\psi\bar{\partial}\psi$ is a chiral primary operator whose conjugate, the anti-chiral primary operator $\psi^{\dagger}\partial\psi^{\dagger}$, appeared in the main text. The anomalous dimension computed above coincides with that expected from the conformal algebra \eqn{acpdim}.

\para
 More generally we can consider the operator $\bar{\partial}^{n}\psi\bar{\partial}^{m}\psi$ with $n\neq m$. For $n+m\geq 3$ these operators suffer polynomial divergence at one-loop which need to be suitably regularized. Nonetheless, their anomalous dimension may be read off from the logarithmically divergent term and can be shown to be
\begin{equation}
\gamma_{\bar{\partial}^{n}\psi\bar{\partial}^{m}\psi}=-\frac{1}{k}
\nn
\end{equation}
This enables us to compute the anomalous dimension of $\tilde{\cO}_n^\dagger=\psi\bar{\partial}\psi\dots\bar{\partial}^{n-1}\psi$, which is the conjugate of the anti-chiral primary operator discussed in the main text \eqn{otilde}. At one loop, the dimension of the operator receives a contribution from each pair $\bar{\partial}^{n}\psi\bar{\partial}^{m}\psi$. But since every pair generates the same contribution $-1/k$ the computation of the anomalous dimension simply reduces to the combinatorial problem of counting all the pairs. The result is
\begin{equation}
\gamma_{\tilde{\cO}_{n}} = -\frac{n(n-1)}{2k}\ \ \ \implies\ \ \ \Delta_{\tilde{\cO}_n} = \frac{n(n+1)}{2} - \frac{n(n-1)}{2k}\nn
\end{equation}
in agreement with the result from the conformal algebra \eqn{acpdim}. Once again, the dimension is one-loop exact.


\begin{thebibliography}{99}

\small
\parskip=0pt plus 2pt

\bibitem{roman}  R.~Jackiw,
  ``{\it Dynamical Symmetry of the Magnetic Vortex},''
  Annals Phys.\  {\bf 201}, 83 (1990).


\bibitem{jpi} R.~Jackiw and S.~Y.~Pi,
  ``{\it Classical and quantal nonrelativistic Chern-Simons theory},''
  Phys.\ Rev.\ D {\bf 42}, 3500 (1990)
  [Phys.\ Rev.\ D {\bf 48}, 3929 (1993)].

\bibitem{son}  Y.~Nishida and D.~T.~Son,
  ``{\it Nonrelativistic conformal field theories},''
  Phys.\ Rev.\ D {\bf 76}, 086004 (2007)
  [arXiv:0706.3746 [hep-th]].

\bibitem{llm}   M.~Leblanc, G.~Lozano and H.~Min,
  ``{\it Extended superconformal Galilean symmetry in Chern-Simons matter systems},''
  Annals Phys.\  {\bf 219}, 328 (1992)
  [hep-th/9206039].
  
  
  \bibitem{nakayama}   Y.~Nakayama,
  ``{\it Index for Non-relativistic Superconformal Field Theories},''
  JHEP {\bf 0810}, 083 (2008)
  [arXiv:0807.3344 [hep-th]].

\bibitem{3lee}  K.~M.~Lee, S.~Lee and S.~Lee,
  ``{\it Nonrelativistic Superconformal M2-Brane Theory},''
  JHEP {\bf 0909}, 030 (2009)
  [arXiv:0902.3857 [hep-th]].
  
\bibitem{karabali}   M.~J.~Bowick, D.~Karabali and L.~C.~R.~Wijewardhana,
  ``{\it Fractional Spin via Canonical Quantization of the O(3) Nonlinear Sigma Model},''
  Nucl.\ Phys.\ B {\bf 271}, 417 (1986).

\bibitem{gold}  A.~S.~Goldhaber and R.~MacKenzie,
  ``{\it Are Cyons Really Anyons?},''
  Phys.\ Lett.\ B {\bf 214}, 471 (1988).
  
\bibitem{djt}   G.~V.~Dunne, R.~Jackiw and C.~A.~Trugenberger,
  ``{\it Chern-Simons Theory in the Schrodinger Representation},''
  Annals Phys.\  {\bf 194}, 197 (1989).
  
\bibitem{jpvort}   R.~Jackiw and S.~Y.~Pi,
  ``{\it Soliton Solutions to the Gauged Nonlinear Schrodinger Equation on the Plane},''
  Phys.\ Rev.\ Lett.\  {\bf 64}, 2969 (1990).
  
\bibitem{nak1} Y.~Nakayama, S.~Ryu, M.~Sakaguchi and K.~Yoshida,
  ``{\it A Family of super Schrodinger invariant Chern-Simons matter systems},''
  JHEP {\bf 0901}, 006 (2009)
  [arXiv:0811.2461 [hep-th]].

\bibitem{nak2}  Y.~Nakayama, M.~Sakaguchi and K.~Yoshida,
  ``{\it Interacting SUSY-singlet matter in non-relativistic Chern-Simons theory},''
  J.\ Phys.\ A {\bf 42}, 195402 (2009)
  [arXiv:0812.1564 [hep-th]].
  
\bibitem{nak3}  Y.~Nakayama, M.~Sakaguchi and K.~Yoshida,
  ``{\it Non-Relativistic M2-brane Gauge Theory and New Superconformal Algebra},''
  JHEP {\bf 0904}, 096 (2009)
  [arXiv:0902.2204 [hep-th]].

\bibitem{murugan}   C.~Lopez-Arcos, J.~Murugan and H.~Nastase,
  ``{\it Nonrelativistic limit of the abelianized ABJM model and the ADS/CMT correspondence},''
  arXiv:1510.01662 [hep-th].
  
\bibitem{seok}  S.~Kim and K.~Madhu,
  ``{\it Aspects of monopole operators in N=6 Chern-Simons theory},''
  JHEP {\bf 0912}, 018 (2009)
  [arXiv:0906.4751 [hep-th]].
  
\bibitem{seok1}   H.~C.~Kim and S.~Kim,
  ``{\it Semi-classical monopole operators in Chern-Simons-matter theories},''
  arXiv:1007.4560 [hep-th].
  
\bibitem{is}   K.~Intriligator and N.~Seiberg,
  ``{\it Aspects of 3d N=2 Chern-Simons-Matter Theories},''
  JHEP {\bf 1307}, 079 (2013)
  [arXiv:1305.1633 [hep-th]].

\bibitem{ofer}  O.~Aharony, P.~Narayan and T.~Sharma,
  ``{\it On monopole operators in supersymmetric Chern-Simons-matter theories},''
  JHEP {\bf 1505}, 117 (2015)
  [arXiv:1502.00945 [hep-th]].


\bibitem{susyqhe}   D.~Tong and C.~Turner,
  ``{\it The Quantum Hall Effect in Supersymmetric Chern-Simons Theories},''
  arXiv:1508.00580 [hep-th].

\bibitem{bergman}   O.~Bergman and G.~Lozano,
  ``{\it Aharonov-Bohm scattering, contact interactions and scale invariance},''
  Annals Phys.\  {\bf 229}, 416 (1994)
  [hep-th/9302116].


\bibitem{bbak}
 D.~Bak and O.~Bergman,
  ``{\it Perturbative analysis of nonAbelian Aharonov-Bohm scattering},''
  Phys.\ Rev.\ D {\bf 51}, 1994 (1995)
  [hep-th/9403104].

\bibitem{lerda}   A.~Lerda,
  ``{\it Anyons: Quantum mechanics of particles with fractional statistics},''
  Lect.\ Notes Phys.\ M {\bf 14}, 1 (1992).
  
\bibitem{henkel}   M.~Henkel,
  ``{\it Schrodinger invariance in strongly anisotropic critical systems},''
  J.\ Statist.\ Phys.\  {\bf 75}, 1023 (1994)
  [hep-th/9310081].
  
\bibitem{aff}  V.~de Alfaro, S.~Fubini and G.~Furlan,
  ``{\it Conformal Invariance in Quantum Mechanics},''
  Nuovo Cim.\ A {\bf 34}, 569 (1976).

\bibitem{nagain} Y.~Nishida and D.~T.~Son,
  ``{\it Unitary Fermi gas, epsilon expansion, and nonrelativistic conformal field theories},''
  Lect.\ Notes Phys.\  {\bf 836}, 233 (2012)
  [arXiv:1004.3597 [cond-mat.quant-gas]].
  
\bibitem{ggt}   J.~P.~Gauntlett, J.~Gomis and P.~K.~Townsend,
  ``{\it Supersymmetry and the physical phase space formulation of spinning particles},''
  Phys.\ Lett.\ B {\bf 248}, 288 (1990).
  
  
\bibitem{duval} C.~Duval and P.~A.~Horvathy,
  ``{\it On Schrodinger superalgebras},''
  J.\ Math.\ Phys.\  {\bf 35}, 2516 (1994)
  [hep-th/0508079].
  

 
\bibitem{chou}  C.~Chou,
  ``{\it The Multi - anyon spectra and wave functions},''
  Phys.\ Rev.\ D {\bf 44}, 2533 (1991)
  [Phys.\ Rev.\ D {\bf 45}, 1433 (1992)].

\bibitem{murthy}   M.~V.~N.~Murthy, J.~Law, R.~K.~Bhaduri and G.~Date,
  ``{\it On a class of noninterpolating solutions of the many anyon problem},''
  J.\ Phys.\ A {\bf 25}, 6163 (1992).




\bibitem{khare}   A.~Khare,
  ``{\it Fractional statistics and quantum theory},''
  Singapore, Singapore: World Scientific (1997) 309 p


 \bibitem{date} G. Date, M.V.N. Murthy, R. Vathsan, ``{\it Classical and Quantum Mechanics of Anyons}", arXiv:cond-mat/0302019


\bibitem{verb0}   M.~Sporre, J.~J.~M.~Verbaarschot and I.~Zahed,
  ``{\it Solution of the three anyon problem},''
  Phys.\ Rev.\ Lett.\  {\bf 67}, 1813 (1991).
  
\bibitem{verb} M.~Sporre, J.~J.~M.~Verbaarschot and I.~Zahed,
  ``{\it Four anyons in a harmonic well},''
  Phys.\ Rev.\ B {\bf 46}, 5738 (1992).

\bibitem{gs}   A.~Giveon and D.~Kutasov,
  ``{\it Seiberg Duality in Chern-Simons Theory},''
  Nucl.\ Phys.\ B {\bf 812}, 1 (2009)
  [arXiv:0808.0360 [hep-th]].

\bibitem{min1}   S.~Giombi, S.~Minwalla, S.~Prakash, S.~P.~Trivedi, S.~R.~Wadia and X.~Yin,
  ``{\it Chern-Simons Theory with Vector Fermion Matter},''
  Eur.\ Phys.\ J.\ C {\bf 72}, 2112 (2012)
  [arXiv:1110.4386 [hep-th]].

\bibitem{ofer1}  O.~Aharony, G.~Gur-Ari and R.~Yacoby,
  ``{\it Correlation Functions of Large N Chern-Simons-Matter Theories and Bosonization in Three Dimensions},''
  JHEP {\bf 1212}, 028 (2012)
  [arXiv:1207.4593 [hep-th]].

\bibitem{min2}   S.~Jain, S.~Minwalla and S.~Yokoyama,
  ``{\it Chern Simons duality with a fundamental boson and fermion},''
  JHEP {\bf 1311}, 037 (2013)
  [arXiv:1305.7235 [hep-th]].
  
   \bibitem{ambak} 
G.~Amelino-Camelia and D.~Bak,
  ``{\it Schrodinger selfadjoint extension and quantum field theory},''
  Phys.\ Lett.\ B {\bf 343}, 231 (1995)
  [hep-th/9406213].
  
\bibitem{amelino} G.~Amelino-Camelia,
  ``{\it Perturbative bosonic end anyon spectra and contact interactions},''
  Phys.\ Lett.\ B {\bf 326}, 282 (1994)
  [hep-th/9402020].

\bibitem{manuel}  C.~Manuel and R.~Tarrach,
  ``{\it Do anyons contact interact?},''
  Phys.\ Lett.\ B {\bf 268}, 222 (1991).
  
  
\bibitem{nishy}   Y.~Nishida,
  ``{\it Impossibility of the Efimov effect for p-wave interactions},''
  Phys.\ Rev.\ A {\bf 86}, 012710 (2012)
  [arXiv:1111.6961 [cond-mat.quant-gas]].
  

\bibitem{beg}   M.~A.~B.~Beg and R.~C.~Furlong,
  ``{\it The $\Lambda \phi^4$ Theory in the Nonrelativistic Limit},''
  Phys.\ Rev.\ D {\bf 31}, 1370 (1985).
  
\bibitem{jbeg} R.~Jackiw,
  ``{\it Delta function potentials in two-dimensional and three-dimensional quantum mechanics},''
  In *Jackiw, R.: Diverse topics in theoretical and mathematical physics* 35-53.
  
\bibitem{holstein} B.~R.~Holstein,
 ``{\it Anomalies for pedestrians},"
 Am.\ J.\ Phys. {\bf 61}, 142 (1992).
  



\bibitem{dunne} G.~V.~Dunne,
  ``{\it Aspects of Chern-Simons theory},''
  hep-th/9902115.
  
 \bibitem{horvathy}   P.~A.~Horvathy and P.~Zhang,
  ``{\it Vortices in (abelian) Chern-Simons gauge theory},''
  Phys.\ Rept.\  {\bf 481}, 83 (2009)
  [arXiv:0811.2094 [hep-th]].
  
\bibitem{jpagain}  R.~Jackiw and S.~Y.~Pi,
  ``{\it Time dependent Chern-Simons solitons and their quantization},''
  Phys.\ Rev.\ D {\bf 44}, 2524 (1991).
  
\bibitem{taubes}   C.~H.~Taubes,
  ``{\it Arbitrary N: Vortex Solutions to the First Order Landau-Ginzburg Equations},''
  Commun.\ Math.\ Phys.\  {\bf 72}, 277 (1980).
  

\bibitem{baklee}  D.~Bak and H.~j.~Lee,
  ``{\it Moduli space dynamics of a first order vortex system},''
  Phys.\ Lett.\ B {\bf 432}, 175 (1998)
  [hep-th/9706102].

\bibitem{htong} A.~Hanany and D.~Tong,
  ``{\it Vortices, instantons and branes},''
  JHEP {\bf 0307}, 037 (2003)
  [hep-th/0306150].





















\end{thebibliography}
\end{document}